\documentclass[a4paper,11pt]{article}

\usepackage{amssymb,amstext,amsmath,amsthm}
\usepackage{graphicx}
\usepackage{latexsym}
\usepackage{psfrag}
\usepackage{amsfonts}
\usepackage{verbatim}
\usepackage{vmargin}
\usepackage{helvet}

\usepackage{amsmath,amsxtra,amsthm,amssymb,xr}
\usepackage[all]{xy}
\usepackage[]{mathrsfs}

\usepackage{color}

\newcommand{\bea}{\begin{eqnarray}}
\newcommand{\eea}{\end{eqnarray}}
\newcommand{\nn}{\nonumber}
\newcommand{\be}{\begin{equation}}
\newcommand{\ee}{\end{equation}}

\newtheorem{theo}{Theorem}
\newtheorem{lem}{Lemma}

\newcommand{\nb}{{\bf n}}
\newcommand{\vb}{{\bf v}}

\newcommand{\Z}{\mathcal{Z}}
\newcommand{\T}{\mathcal{T}}

\newcommand{\R}{\mathbb{R}}
\newcommand{\C}{\mathbb{C}}
\newcommand{\imm}{\mathfrak{i}}

\def\bra#1{\mathinner{\langle{#1}|}}
\def\ket#1{\mathinner{|{#1}\rangle}}
\def\braket#1#2{\mathinner{\langle{#1}|{#2}\rangle}}

\newcommand{\SU}{\mathrm{SU}}
\newcommand{\SO}{\mathrm{SO}}
\newcommand{\OO}{\mathrm{O}}

\newcommand{\U}{\mathrm{U}}

\newcommand{\Reel}{\mathrm{Re}}

\newcommand{\pole}{\mathcal N}

\newcommand{\dd}{\mathrm{d}}
\newcommand{\torus}{\mathbb{T}}

\def\la{\langle}
\def\ra{\rangle}

\setmarginsrb{2.5cm}{1cm}{2.5cm}{2cm}{12pt}{11mm}{0pt}{13mm}
\title{Asymptotic analysis of the Ponzano-Regge model for handlebodies}
\author{Richard J. Dowdall\footnote{richard.dowdall@maths.nottingham.ac.uk}
,  Henrique Gomes\footnote{henrique.gomes@maths.nottingham.ac.uk}
, Frank Hellmann\footnote{frank.hellmann@maths.nottingham.ac.uk}
\vspace{3mm}
\\ School of Mathematical Sciences \\ Nottingham University \\ University Park \\ Nottingham NG7 2RD \\ UK}

\begin{document}

\maketitle

\vspace{-3mm}
\begin{abstract}
Using the coherent state techniques developed for the analysis of the EPRL model we give the asymptotic formula for the Ponzano-Regge model amplitude for non-tardis triangulations of handlebodies in the limit of large boundary spins.
The formula produces a sum over all possible immersions of the boundary triangulation and its value is given by the cosine of the Regge action evaluated on these. Furthermore the asymptotic scaling registers the existence of flexible immersions.
We verify numerically that this formula approximates the 6j-symbol for large spins.
\end{abstract}

\begin{comment}
\section{Conventions}

$$g \in \SU(2)$$

$$\hat{g} \in \SO(3)$$

$\la Tet \ra$ is a spin network evaluation. Tet, Theta are the spin network graphs.\\

$\Sigma$ is a three manifold.\\

$L^i$ are the $\SU(2)$ generators with $[L^i,L^j] = - \epsilon^{ijk} L^k$  and
therefore $L^i = i/2 \sigma^i$ (check that)\\
\end{comment}

\section{Introduction}
%%%%%%%%%%%%%%%%%%%%%%%%

In \cite{ponzanoregge}, Ponzano and Regge gave a formula  for the large spin limit of the 6j symbol.  It was found
to be related to the Regge action for discrete general relativity and with this motivation they constructed the
first spin foam model of 3d gravity. Their asymptotic formula was first proved in \cite{roberts-1999-3} then more
recently using different methods in \cite{gurau-2008,garoufalidis-2009} and the square of the 6j symbol was also studied in the
context of relativistic spin networks \cite{barrett-2003-20,Freidel:2002mj}.  The next to leading order
approximation was recently considered in \cite{Dupuis:2009sz}. The precise formulation of the full state sum was studied in \cite{Barrett:2008wh}.

In \cite{Barrett:2009gg,Barrett:2009mw}, the semiclassical limit of some  recent spin foam models
\cite{Freidel:2007py,Engle:2007wy,livine-2007-76} was analysed using the coherent state techniques introduced in
\cite{livine-2007-76}. In particular the boundary there was formulated in term of coherent tetrahedra.   Here we
apply the same techniques in the 3d case using coherent triangles and, instead of a single vertex amplitude, we
analyze triangulations of arbitrary genus handlebodies.  This finally opens up the possibility of a
continuum limit and renormalization analysis of the model for this restricted class of 3-manifolds.  In particular the resulting formula is well suited for studying the graviton propagator as introduced for Ponzano-Regge in \cite{Speziale:2005ma}.

We begin the paper by describing the formulation of the Ponzano-Regge model in terms of a  single spin network
diagram dual to the boundary triangulation. We then describe the boundary state choice in detail and proceed to
give the asymptotic formula in terms of immersions of the boundary triangulation. The asymptotic scaling of the
amplitude has the interesting feature that it registers whether or not there are flexible immersions of the
boundary. An explicit example is provided in Steffen's polyhedron. This analysis sheds light on the general way in
which asymptotic behaviour for larger triangulations can emerge from a spin foam model. In particular it does not
need to proceed by taking the asymptotics of the individual simplex amplitudes first.

\section{The Ponzano-Regge model in terms of coherent states on the boundary}
%%%%%%%%%%%%%%%%%%%%%%%%

The Ponzano Regge amplitude was originally defined in terms of 6j symbols with a cutoff regularization on the interior vertices.
More recently, it was shown in \cite{Barrett:2008wh} that the cutoff regularization for sums over
representations in some cases disallows a 2-3 Pachner move (the Biedenharn-Elliot identity does not hold for a restricted
sum over representations.) This meant topological invariance of the partition function can not be proved with Pachner moves in this form.
The alternative formulation in terms of delta functions and integrals over $\SU(2)$ regularized with a gauge fixing tree is both finite and invariant under Pachner moves.  Another regularization using representations of a quantum group is given by the Turaev-Viro model, however it was also shown in \cite{Barrett:2008wh} that the limiting procedure that gives the Ponzano-Regge model is only known to exist for so-called non-tardis triangulations - i.e. a triangulation whose edge lengths are restricted to a finite range by the boundary edge lengths.
In order to avoid discussing regularization, in this paper we will restrict to only considering these non-tardis triangulations which are by definition finite.
 Slightly extending the terminology of \cite{Barrett:2008wh}, we will call a manifold $\Sigma^3$ a {\it non-tardis manifold} if there exists a `non-tardis' triangulation of $\Sigma^3$.

 For a 3-manifold $ \Sigma^3$ with orientable 2-boundary $\partial \Sigma^3$ its boundary state space is then given by the possible geometric triangulations of the 2-boundary with half integer edge lengths. The amplitude for such a non-tardis manifold is given in terms of a
non-tardis triangulation $\T$ of $\Sigma^3$ that extends the boundary triangulation, and some boundary state
$\Psi$:
\be
\label{PR-formal}
\Z_{PR}(\Psi, \T) = \sum_{j_e} \prod_{e} \dim(j_e) \prod_{\Delta}\frac{1}{\la {\rm
Theta} \ra} \prod_{\sigma} \la {\rm Tet}\ra.
\ee
Here $e$ is an edge, $\Delta$ a triangle and $\sigma$ a
tetrahedron of the triangulation of the interior,  $j$ are half integers labelling the irreps of $\SU(2)$. The
amplitudes $\la {\rm Theta} \ra $ and $\la {\rm Tet} \ra$ are the spin network evaluation of the theta graph and
the planar tetrahedral spin network respectively. These spin networks are the two dimensional duals to the
interior triangles $\Delta$ and, respectively, to the surface of the tetrahedra $\sigma$ in the triangulation. The
labelling of the spin network surface duals of the $\Delta$ and $\sigma$ is given by assigning the $j$ associated
to each edge to each dual edge that crosses it. Finally $\dim(j) = (-1)^{2j} (2j+1)$ is the (super)-dimension of
the $j$th $\SU(2)$ irrep in graphical calculus.

In the interior the normalisation and phase of the intertwiners cancels. However, at the boundary these are arbitrary normalisation for each face. This information is in the boundary state $\Psi$ which consists of the boundary edge length data and the particular intertwiner chosen at each face.

\subsection{Ponzano Regge on the boundary}

In some cases it is possible to reformulate the  Ponzano-Regge model defined above as a spin
network evaluation on the 2-boundary of the manifold.
In fact, Ponzano and Regge originally constructed the state sum model such that it agreed with the evaluation of a planar spin network associated to the boundary of a 3-ball.  An algorithm to construct a non-tardis interior triangulation given an arbitrary triangulation of the boundary of
$B^3$, was given using recoupling theory by Moussouris in \cite{Moussouris}.  This algorithm consists of reducing the boundary spin network to a product of 6j symbols (which is always possible for a planar diagram) using the recoupling identity and Schur's Lemma and then reconstructing the interior triangulation from these 6j symbols.
Since the boundary spin network is finite, this procedure gives a manifestly finite definition of the partition function.
%that is explicitly independent of the interior of the manifold.
%As this formulation is the one we analyze in this paper we will describe in some detail how it can be obtained from \eqref{PR-formal}.

%In this paper we will extend this result to spin networks on the boundary of handlebodies of arbitrary genus.
%A non-tardis triangulation of a handlebody of genus $g$ can be constructed as follows.
%Start with a triangulation of the boundary of $B^3$. A non-tardis triangulation of the interior of the ball is given by applying the Moussouris %algorithm and a triangulation of the handlebody can be formed by identifying $g$ distinct pairs of triangles.  This identification must not identify %any triangles that share a vertex or whose vertices are in a common tetrahedron.

In this paper we will extend this result to spin networks on the boundary of handlebodies of arbitrary genus.
A non-tardis triangulation of a handlebody of genus $g$ can be constructed as follows. Start with a triangulation of the boundary of $B^3$ with $g$ distinct pairs of triangles that do not share a common vertex.  The boundary of the handlebody can be formed by identifying these triangles and a non-tardis triangulation of the interior is given by applying the Moussouris algorithm.  This procedure may result in a degenerate triangulation of the handlebody even if the triangulation of the ball is non-degenerate.

We will begin our analysis by reformulating the amplitude for the 3-ball $B^3$ on a non-tardis triangulation as a spin network on the boundary by a ``reverse Moussouris algorithm.'' We then describe how this procedure is altered for handlebodies  of arbitrary genus.
From now on, $\Sigma^3$ denotes a handlebody.

\begin{lem}
The Ponzano-Regge amplitude for a non-tardis triangulation of $B^3$ can be expressed in terms of a single spin network evaluation
\be
\Z_{PR}(\Psi, B^3) = \la (\partial B^3)^* \ra
\ee
where $^*$ is the two dimensional dual of the surface
triangulation with each  dual edge labelled with the $\SU(2)$ irrep corresponding to the length of the edge it is
dual to, and the spin network is evaluated as the planar projection without crossings, with the intertwiner
normalisation given by $\Psi$.

\end{lem}
{\em Proof}: In order to reexpress the 3-ball with a given triangulation $\mathcal{T}$ and the  amplitude $\Z(\Psi,B^3)$ as the spin
network evaluation of its boundary we proceed inductively. Note first that a triangulation of $B^3$ given by a
single tetrahedron is already of the form we want to put it in: by \eqref{PR-formal} its amplitude is given
exactly by the evaluation of the spin network dual to its boundary 2-geometry with an intertwiner normalisation
chosen at each surface triangle. This establishes the base case. We now need to show that the statement remains true when one glues tetrahedra on to a ball amplitude already expressed in this manner, and thus reconstruct arbitrary non-tardis triangulations
of the 3-ball. To glue we add the necessary face and edge amplitudes for the new interior faces and edges with the same the normalisation and phase choice of the intertwiners chosen in the boundary state before. These boundary choices will therefore cancel. This is in accordance with the observation above that the phase choice and normalisation on the interior are left arbitrary. A tetrahedron can be glued onto a ball non-degenerately with one or two faces:

\begin{enumerate}
    \item If we glue one face of the tetrahedron with one face
    of the ball we create an inner triangle. The PR amplitude  of the new ball differs from the old one by a
    $\frac{1}{\la {\rm Theta} \ra}$ and a tetrahedral net. In the spin network evaluation the vertices of the
    3-ball and the tetrahedral amplitude corresponding to the glued face, together with the face amplitude,
    are the normalized projector on the invariant subspace of the irreps on the edges. As both amplitudes
    being glued already are invariant we can simply replace them with parallel strands, see Figure \ref{ball
    gluing fig}. This changes the spin network graph being evaluated by changing a vertex to a triangle. This
    is the dual to the change of the surface triangulation, and the resulting amplitude still satisfies the
    lemma.

    \item If we glue two faces of the tetrahedron onto the ball,  we create an inner edge  and two inner faces,
    the PR amplitude changes by adding a tetrahedral net, two thetas and one dimension factor. However, these
    nets correspond exactly to the $6j$ symbol for changing the ball amplitude from being connected along the
    dual of the old boundary to the dual of the new boundary.
    %\item If we glue three faces of the tetrahedron onto the ball we create three inner edges, three inner faces and
    %one internal vertex. A small calculation shows that the resulting three dimension factors and theta
    %networks, together with the tetrahedral network can be rewritten as a single vertex up to an infinite
    %constant $\sum_j \dim_j^2$ which cancels the infinite constant normalization corresponding to the inner
    %vertex.
\end{enumerate}
\begin{figure}
\begin{center}
\begin{tabular}{c}
  \psfrag{j1}{$j_1$}
\psfrag{j2}{$j_2$}
\psfrag{j3}{$j_3$}
%\psfrag{j4}{$j_4$}
%\psfrag{j5}{$j_5$}
%\psfrag{j6}{$j_6$}
%\psfrag{j7}{$j_7$}
%\psfrag{j8}{$j_8$}
%\psfrag{j9}{$j_9$}
  \includegraphics[scale=0.25]{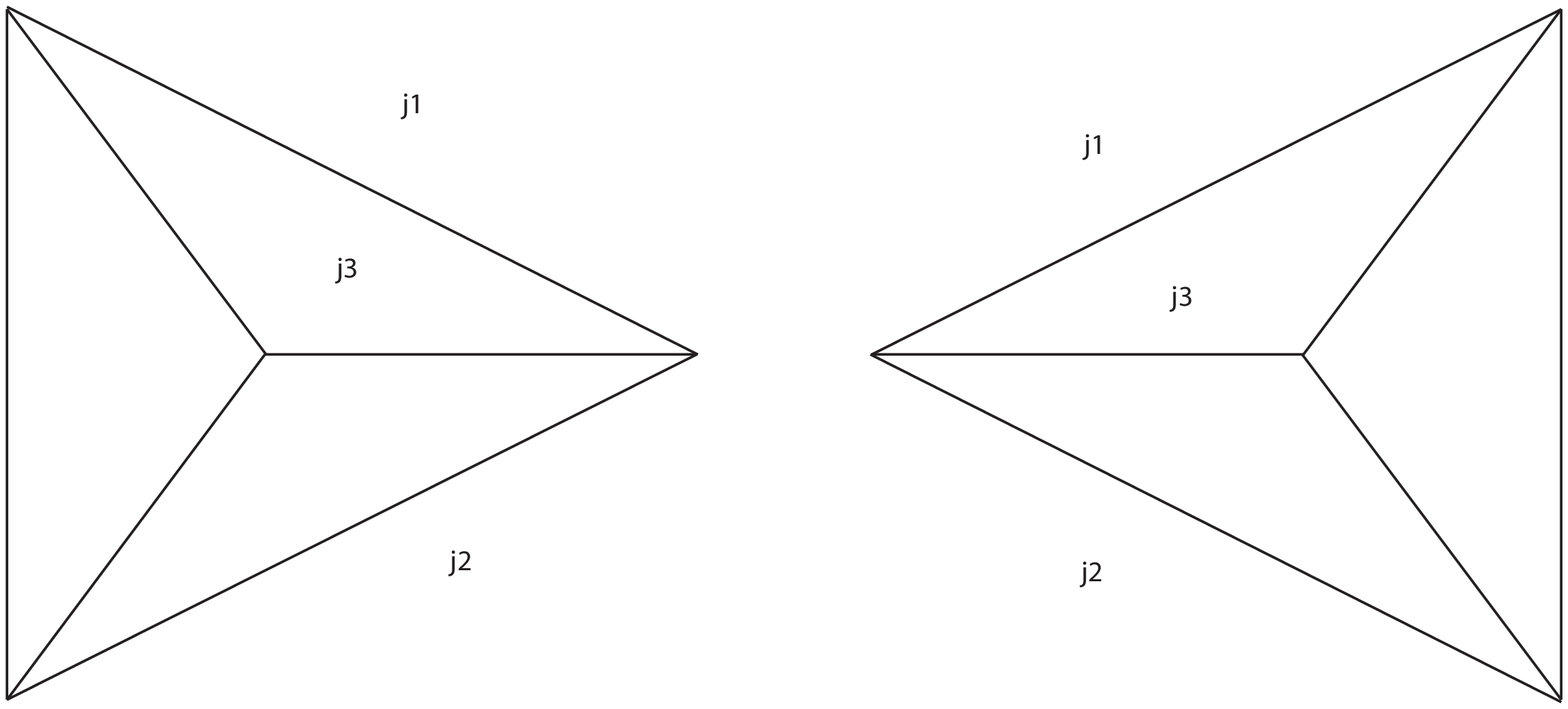}
\hspace{0.2cm}
\end{tabular}
$
\begin{array}{c}
  = \theta(j_1,j_2,j_3)
\end{array}
$
\begin{tabular}{c}
\hspace{0.2cm}
  \psfrag{j1}{$j_1$}
\psfrag{j2}{$j_2$}
\psfrag{j3}{$j_3$}
%\psfrag{j4}{$j_4$}
%\psfrag{j5}{$j_5$}
%\psfrag{j6}{$j_6$}
%\psfrag{j7}{$j_7$}
%\psfrag{j8}{$j_8$}
%\psfrag{j9}{$j_9$}
\includegraphics[scale=0.25]{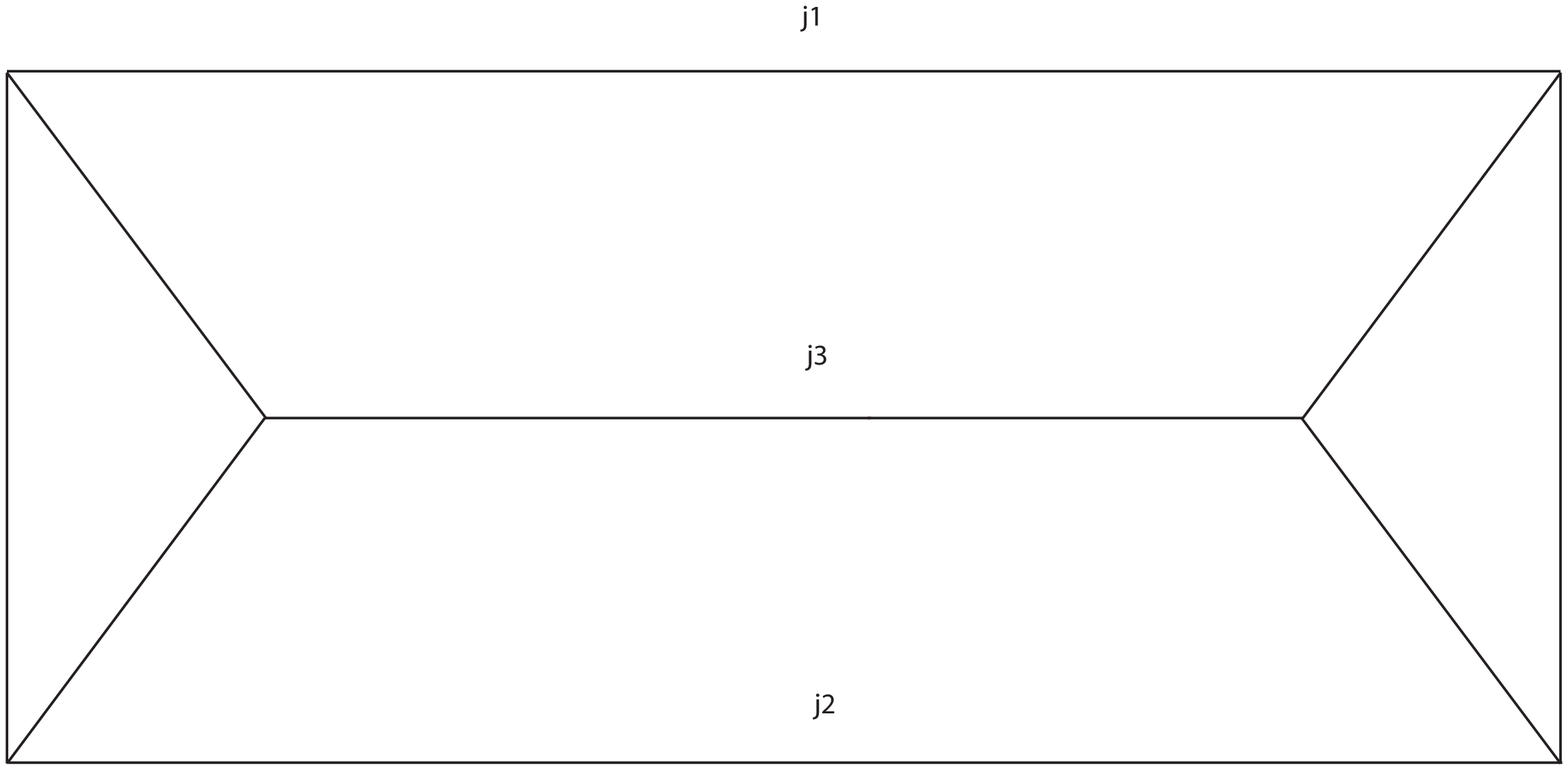} \\
\end{tabular}

\caption{Case 1: Reduction of the PR amplitude for two tetrahedra to the  spin network on the boundary.
}\label{ball gluing fig}
\end{center}
\end{figure}
Note that for any non-tardis triangulation of $B^3$ we can always build it up from a single tetrahedron by gluing on one or two faces. Furthermore the two operations described above do not introduce crossings and respect the planar projection chosen.

This establishes that one can express the Ponzano Regge amplitude of an  arbitrary triangulation of $B^3$ as a
spin network evaluation on its boundary $S^2$. This proves the lemma $\square$.\smallskip

Consider next the case of a solid torus $D^2\times S^1$, which we call
$\torus$. Take a disc $\mathbb{D}\subset\torus$ such that $\mathbb{D} = D^2\times\{p\}\in D^2\times S^1$. For future purposes, note that it intersects a non-contractible loop in $\torus$. We can now always move this disc by a homotopy that keeps $\partial \mathbb{D}$ on $\partial \torus$ such that its boundary is the union of at least three edges of the boundary triangulation. Due to triangulation invariance
we can then choose a triangulation such that $\mathbb{D}$ has no internal vertex. We can then cut the Ponzano Regge
amplitude along this surface, the resulting space is topologically $B^3$ and we can apply the previous lemma. This
yields a ball where two discs on the boundary are glued by identifying edges and using the PR face and edge
weights. Call $n$ the number of edges that make up $\partial\mathbb{D}$. As we chose a disc with no internal vertex, the
spin network dual to it has to be an $n-2$ vertex string with one outgoing edge per inner vertex, and two at the
end. Together with the face amplitudes this is simply the projector onto the invariant subspace of the irreps on
the circle $\partial\mathbb{D}$. This projection can then be replaced by a group averaging on the strands dual to
$\partial\mathbb{D}$:

\begin{figure}
\begin{center}
%,trim=0mm -10mm -40mm -10mm
\psfrag{h}{$ h $}
\includegraphics[scale=0.30]{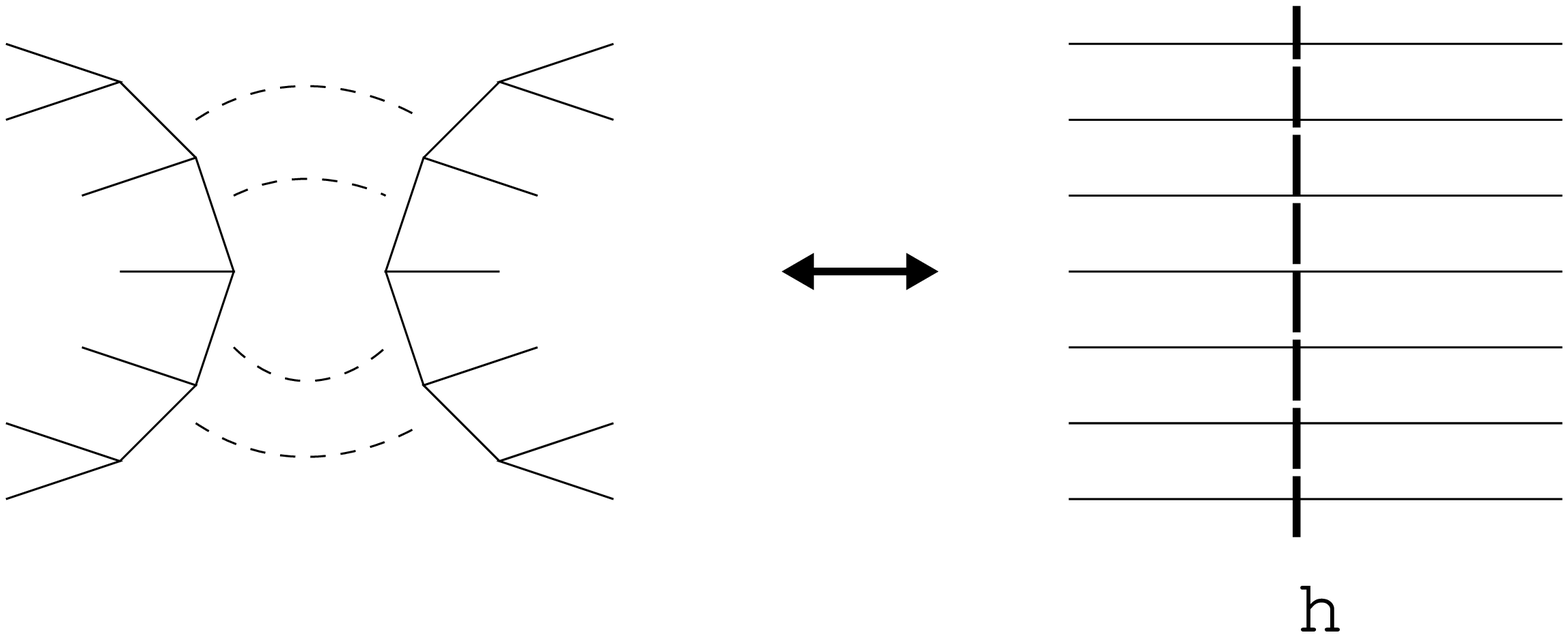}\hspace{1cm}
\includegraphics[scale=0.25]{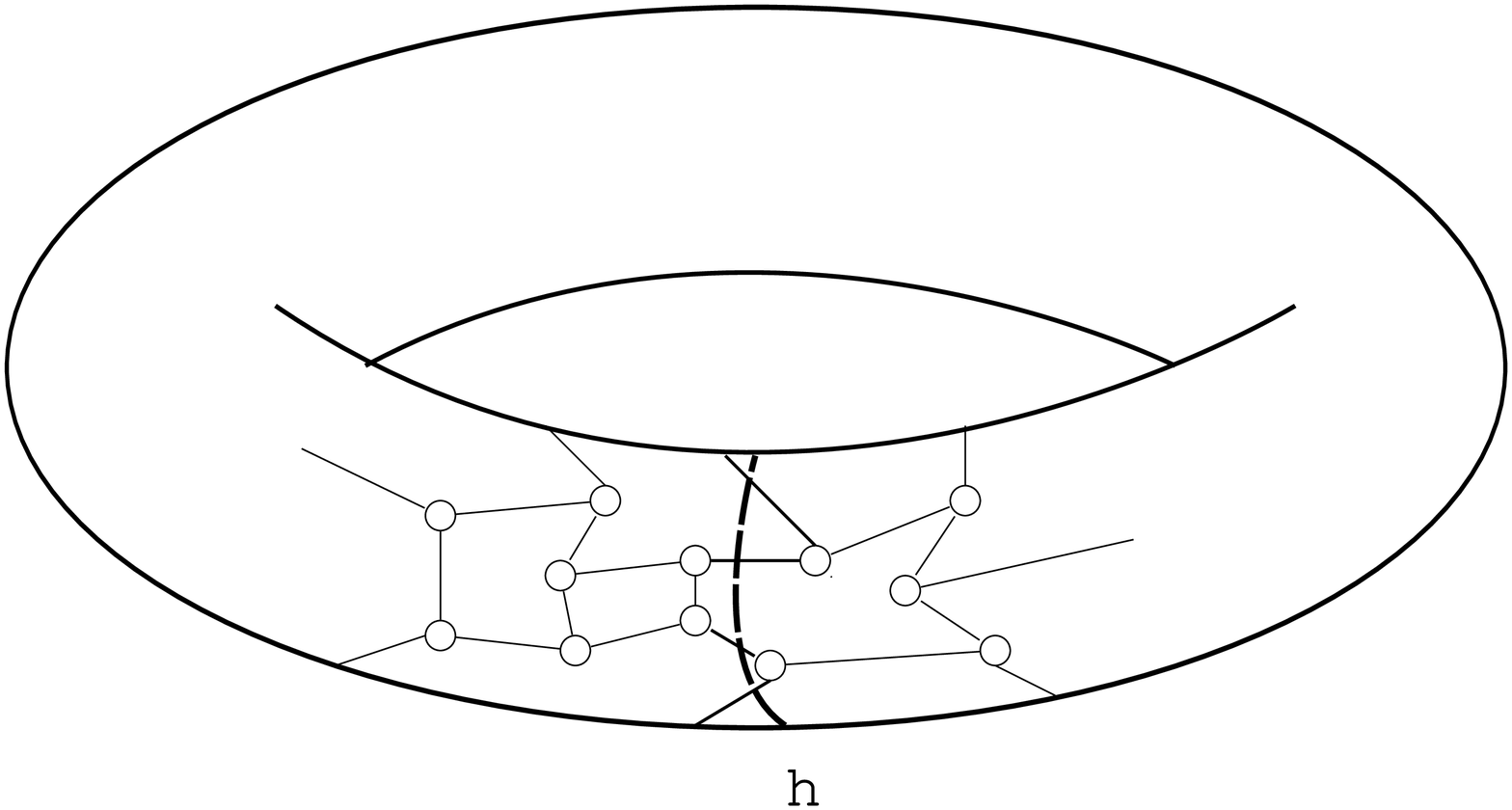}
\caption{By replacing the gluing along the 4 faces with a group averaging on edges crossing the dashed line
(left) we reexpress the $\torus$ amplitude on the boundary (right).}\label{figone}
\end{center}
\end{figure}

\be \Z_{PR}(\Psi, \torus) = \int_{\SU(2)} \dd h \la (\partial \torus)^*_h \ra \ee where $(\partial \torus)^*_h$ is
the spin network dual to the surface  of the torus with $h$ inserted along dual edges crossing
$\partial\mathbb{D}$. The diagram is defined by first cutting along this circle, choosing the planar no crossing
diagram of the graph and then connecting up along the identified edges. See figure \ref{figone}, and Appendix \ref{tet appendix} for an explicit example.

We can easily generalize this example to arbitrary genus handlebodies.
By definition, a handlebody of genus $g$ comes equipped with a set of $g$ standard cuts that reduce the handlebody to the 3-ball.  We call these cuts $\mathbb{D}_i$ with an index $i \in C$ where $C$ is a set of labels for the standard cuts.
For later use, note that it is always possible to define a complete set of generators $c_i$ of the homology group $H_1(\Sigma^3)$ such that each $c_i$ is transversal to the cut $\mathbb{D}_i$ and does not intersect the other cuts.
We can choose an equivalent set of cuts that are related to the standard cuts by boundary preserving homotopy  as long as the cuts remain non-intersecting.
In particular from now on we will choose the cuts so as to lie on the triangulation. This implies a restriction on the class of triangulations considered as such a choice may not exist for small triangulations.

 Now we can state:

\begin{lem}
\label{lemma2}
\be
\Z_{PR}(\Psi, \Sigma^3) = \int_{\SU(2)} \prod_{i\in C}\dd h_i \la (\partial \Sigma^3)^*_{h_i} \ra
\ee
where
$\Sigma^3$ is a handlebody,
$\partial \Sigma^3$ is its triangulated boundary which carries half integer labels on its edges and  $C$ labels the cuts. Choose a set of cuts $\mathbb{D}_i$ that lie on the triangulation.
$ \la (\partial \Sigma^3)^*_{h_i} \ra$ is the spin network evaluation of the dual of the triangulation of the surface, with the links labelled by the half integer lengths of the edges
they cross and a $h_i \in \SU(2)$ inserted on every link that crosses a cut $\partial\mathbb{D}_i\in\partial\Sigma^3,~ i\in C$. The spin network is evaluated in the planar projection of the boundary of the cut manifold. That is, with all crossings being due to the links crossing a cut.
\end{lem}

Proof: Cutting $\Sigma^3$ along the discs $\mathbb{D}_i$ reduces it to a 3-ball.
The spin network evaluation is defined by taking
the planar no crossing representation of the graph cut along the circles $\partial\mathbb{D}_i$ and then
connecting the identified open ends. If we choose a triangulation that triangulates each disc $\mathbb{D}_i$
without internal vertices and reexpress the resulting amplitude as a spin network evaluation, then the gluing of
the faces corresponds to a projection onto invariant subspaces. Replace the projection onto the invariant subspace
by a group integration and we get the lemma. $\square$

Note that due to the intertwining property of the spin network the choice of $\mathbb{D}_i$ does not matter, it merely moves the $h_i$ insertion in the intertwiner around.

\subsection{Coherent triangles}

In order to have a clear geometric picture of the amplitude we will choose the intertwiners in the boundary state
$\Psi$ by using coherent states $\alpha_k(\nb,\theta)$ \cite{perelomov}. These are the highest weight eigenstates of the normalized Lie
algebra elements, that is for $L^i = \frac{i}{2}\sigma^i$ the Lie algebra generators and $\nb \in S^2$, a coherent
state $\alpha_k(\nb,\theta)$ in the $k$ representation satisfies:
\be L.\nb \alpha_k(\nb,\theta) = i k \alpha_k(\nb,\theta)
\ee
The parameter $\theta$ describes a choice of representative of the $\U(1)$ equivalence class of states that correspond to the same $\nb$.
These states transform with a  phase under the group elements generated by $L.\nb$ and the label $\nb$ transforms
covariantly under the $\SO(3)$ action of $\SU(2)$. That is for $g\in \SU(2)$ with corresponding $\SO(3)$ element $\hat{g}$:
\be
\label{act on coherent} g
\alpha_k(\nb,\theta) = e^{i k \phi} \alpha_k(\hat{g}\nb,\theta)
\ee
For the asymptotic analysis three further properties will be
crucial:

\begin{itemize}
  \item The $k$ representation can be constructed as the symmetric  subspace of $2k$ copies of the fundamental
  representation. In this picture coherent states decompose into a tensor product of coherent states in the
  fundamental representation. Consequently the group action factorizes:
  \be
  g \alpha_k(\nb,\theta) = g \bigotimes_{i=1}^{2k}  \alpha_{\frac{1}{2}}(\nb,\theta) = \bigotimes_{i=1}^{2k}  e^{i \frac{\phi}{2}} \alpha_{\frac{1}{2}}(\hat{g}\nb,\theta).
  \ee
  \item The modulus squared of the Hermitian inner product of coherent states is given by:
  \be
  \label{coh state mod sq}
  |\langle   \alpha_k(\nb_1,\theta_1) , \alpha_k(\nb_2,\theta_2) \rangle  |^2 = \left( \frac{1}{2} (1 + \nb_1 . \nb_2) \right)^{2k},
  \ee
  \item Under the action of the
  standard antilinear structure on $\SU(2)$ (see \cite{Barrett:2009gg}) the coherent state changes as:
  \be L.\nb J
  \alpha_k(\nb,\theta) = -i k  J \alpha_k(\nb,\theta)
  \ee
  The antilinear map $J$ is given by multiplication by the epsilon tensor in the spin $k$ representation followed by complex conjugation.  $J$ commutes with $\SU(2)$ elements.
\end{itemize}

Note that given a set of three edge  labels $k_i$ there is a non zero intertwiner exactly if they satisfy the
triangle inequalities. Therefore there is a set of ${\nb_i}$, unique up to $\OO(3)$ such that $\sum_{i=1}^3 k_i
\nb_i = 0$. We can then choose our intertwiner in the boundary state $\Psi$ as
\be
\iota
= \int_{\SU(2)} \dd X
\bigotimes_i (X \alpha_k(\nb_i,\theta_i))
\ee
This state is clearly an $\SU(2)$ invariant state. As we noted that $\SU(2)$ acts
covariantly as $\SO(3)$ on the labels $\nb_i$ this choice is only dependent on
%the remaining parity $P =\OO(3)/\SO(3)$, and
an unspecified phase as we left open which eigenstates of $L.\nb_i$ we are using. In particular it does not depend on the remaining parity $P =\OO(3)/\SO(3)$ as this acts on the plane of the triangle as an $\SO(3)$ element.

Thus choosing normalized $\alpha_k(\nb,\theta)$  compatible with the boundary spin labels fixes the intertwiner states up
to a parity choice and up to a phase. These two data will be fixed by considering the gluing of the boundary.

\subsection{Regge state}

Let $V$ be a set of labels for the boundary faces. Then we label the boundary edges by pairs $ab ~|~ a,b \in V$
and call the set of such pairs $E$. Let $\phi_a:\Delta_a\rightarrow\pole^\bot$ be an orientation preserving map
from the $a$-th triangle on $\partial\Sigma^3$  to the plane orthogonal to the north pole of $S^2$ (which we
denote $\pole = (0,0,1)$). We choose the orientation in $\pole^\bot$ to be the one inherited from $\mathbb{R}^3$
by taking $\pole$ to be the outward surface normal. As the boundary of $\Sigma^3$ is orientable, we can define
$\nb_{ab}=\phi_a(e_{ab})$ where $ab \in E$.

The requirement that $\phi_a$ be orientation preserving implies that the triangles with the edge vectors given by
$k_{i} \nb_{i}$ all have the same orientation in $\pole^\bot$. In particular we can require them to have the same
orientation as we have chosen for $\pole^\bot$.
%This fixes the parity ambiguity in the coherent intertwiner above.
 In particular this implies
that we can glue up any two triangles $a$,$b$ with a common edge in $\pole^\bot$ in an orientation preserving way.

Thus there exists an element $\hat{g}_{ab} \in \SO(3)$ such that:
\bea
- \nb_{ba} &=& \hat{g}_{ab} \nb_{ab}
\nn\\
\pole &=& \hat{g}_{ab} \pole
\eea
Where  $\nb_{ab}$ is the edge vector of triangle $a$ that  gets glued to
triangle $b$. As in \cite{Barrett:2009gg}, $\hat{g}_{ab}$ is the Levi-Civita parallel translation from triangle $a$ to
triangle $b$, according to the bases provided by $\phi_a$ and $\phi_b$. Again, given a choice of spin structure
for $\Sigma^3$, a choice of a spin frame for each triangle defines the $\SU (2)$ lift $g_{ab}$ as the parallel
translation of the spin connection in these frames.

Next we will describe a canonical choice of phase for the boundary state $\Psi$. From \eqref{act on coherent},
$\alpha_k(- \nb_{ab},\theta_{ab})$ is proportional to $g_{ab} \alpha_k(\nb_{ba},\theta_{ba})$.  Then we fix the relative phase of the
coherent states on the boundary:
\be
J \alpha_k(\nb_{ba},\theta_{ba}) = g_{ab} \alpha_k(\nb_{ab},\theta_{ab})
 \label{costateglue}
\ee
We call coherent states with the above relative state choice Regge states, and denote them $\ket{\nb,k}$
Their
image under the antilinear structure is $\ket{-\nb,k} = J \ket{\nb,k}$, and states in the fundamental representation are denoted $|\nb \rangle$.

The total boundary state is  then given by:
\be
\Psi(k_i , \nb_i) =
\int{ \left(\prod_{a \in V} \dd X_a\right) \bigotimes_{cd
\in E} X_c \; \ket{\nb_{cd},k_{cd}}}
\ee
Due to the presence of the antilinear map in the definition of the
relative phase the overall ambiguity not fixed by \eqref{costateglue} cancels in the overall state. At each
triangle $a$ we have a sign freedom as adding a sign contributes $(-1)^{2 \sum_{b, ab \in E} k_{ab}} = 1$ by the
admissibility conditions on intertwiners. This shows that as  in \cite{Barrett:2009gg} the possible lifts of
$\hat{g}_{ab}$ are defined by the spin structures on the boundary, and do not rely on the arbitrary spin frame
covering chosen to define the lift.

Finally note that inverting the  orientation of $\pole^\bot$ would  have the same effect as turning the state into
$\Psi' = J \Psi$.

\subsection{The Amplitude}

We begin with $B^3$.
To evaluate the spin network defining our amplitude in terms of these coherent  intertwiners we choose a
particular diagrammatic representation of the planar graph. To obtain the spin network evaluation of this graph we
then contract the intertwiners chosen using the epsilon inner product defined in terms of the Hermitian inner
product by $(\alpha,\beta) = \braket{J\alpha}{\beta}$. Number the triangles in the graph from left to right. We
then assume that the coherent intertwiners have been specified with respect to this planar representation of the
graph as well. Then we have no crossings in the diagram and we can now explicitly write the contraction of
coherent intertwiners as: \bea \Z_{PR}(\Psi, B^3)
&=& \int \prod_{a \in V} \dd X_a \prod_{bc \in E} (X_b \ket{\nb_{bc}, k_{bc}},X_c \ket{\nb_{cb},k_{bc}}) \nn\\
%&=& \int \prod_{a \in V} \dd X_a \prod_{bc \in E} \bra{\nb_{bc},k_{bc}}J X_b^\dagger X_c \ket{\nb_{cb}, k_{bc}} \nn\\
&=& \int \prod_{a \in V} \dd X_a \prod_{bc \in E} \bra{- \nb_{bc}, k_{bc}} X_b^\dagger X_c \ket{\nb_{cb},k_{cb}}\nn\\
&=& \int \prod_{a \in V} \dd X_a \prod_{bc \in E} \bra{- \nb_{bc}} X_b^\dagger X_c \ket{\nb_{cb}}^{2k_{bc}} \eea
Where we have written $\ket{\nb_{cb}}$ for $\ket{\nb_{cb},\frac{1}{2}}$.

For general manifolds we need to make sure that, after we have chosen the circles, the dual edges crossing a
circle all have the same orientation relative to the circle. This can be done by using a planar representation
that has all the glued discs strictly left or right of each other. Call $\tilde{E}$ the set of edges not crossing
circles and $E_j$ the set of edges crossing circle $j \in C$. The amplitude is then given by
\be
\Z_{PR}(\Psi,
\Sigma^3)
=
(-1)^\chi \int \prod_{a \in V} \dd X_a \prod_{j \in C} \dd h_j \prod_{bc \in \tilde{E}} \bra{-
\nb_{bc}} X_b^\dagger X_c \ket{\nb_{cb}}^{2k_{bc}} \prod_{l \in C} \prod_{de \in E_l} \bra{- \nb_{de}} X_d^\dagger
h_l X_e \ket{\nb_{ed}}^{2k_{de}}
\label{TheAmplitude}
\ee Where $(-1)^\chi$ is a sign factor incurred in the spin
network evaluation when connecting up the glued edges in the spin network evaluation. This can then be written as
\be
\Z_{PR}(\Psi,\Sigma^3) = (-1)^\chi \int \prod_{i \in V} dX_i \prod_{j \in C} dh_j e^S
\ee
with the action
given by
\be
\label{action} S = \sum_{ab \in \tilde{E}} 2k_{ab}  \ln  \bra{{\bf n}_{ab}}J X_{a}^{\dagger} X_{b}
\ket{{\bf n}_{ba}} + \sum_{l \in C}{\sum_{de \in E_l} 2k_{de} \ln \bra{{\bf n}_{de}}J X_{d}^{\dagger} h_l X_{e}
\ket{{\bf n}_{ed}}}.
\ee
Note that the ambiguity in the logarithm of a complex number does not affect the
amplitude.

\subsection{Symmetries of the action}\label{Symmetries of the action}
The action \eqref{action} has the following symmetries (up to $2 \pi i$)
\begin{itemize}
  \item {\em Continuous.} A global rotation $Y \in \SU(2)$ acting on each $X_a$ and $h_i$ as $X_a \rightarrow Y X_a$
  and $h_i \rightarrow Y h_i Y^{-1}$. This represents a rigid motion of the whole manifold.
  \item {\em Discrete.}  At each triangle $a$ the transformation $X_a \rightarrow \epsilon_a X_a$ with $\epsilon_a = \pm 1$
  leaves a factor $\epsilon_a^{\sum_{b ,ab \in E} 2k_{ab} }$. As the admissibility conditions are satisfied on
  each triangle, this factor equals one. Similarly we have an arbitrary sign $\epsilon_i$ on $h_i$ as
  the edges on which $h_i$ act satisfy the admissibility condition for intertwiners.
\end{itemize}
This latter symmetry will be used to compensate for the ambiguity of the lifts  of $\SO(3)$ to
  $\SU(2)$. We can now state the theorem on the asymptotic formula.

\subsection{Relation to the standard intertwiner phase choice}

The standard choice of phase for an intertwiner, defined by chromatic evaluation \cite{KauffmanLins}, gives real numbers for a spin network evaluation.
We will now show that with the Regge phase choice the amplitude is real so can only differ from the chromatic evaluation by $\pm1$ and a normalisation factor.
Note that since the Regge choice has all the $\nb_{ab}$ orthogonal to ${\bf e}_z$, the rotation $e^{-i\pi{\bf e}_z\cdot\sigma}$ rotates $\nb_{cb}$ to $-\nb_{cb}$ and leaves ${\bf e}_z$ invariant. Under this rotation, the coherent state $\ket{\nb_{cb}}$ will transform as
\be
e^{-i\pi{\bf e}_z\cdot\sigma}\ket{\nb_{cb}} = e^{i\phi}J\ket{\nb_{cb}}
\ee
for some phase $\phi$.

Consider a single term in the amplitude \eqref{TheAmplitude}, and rewrite it
inserting the identity:
 \bea
  \bra{- \nb_{bc}} X_b^\dagger
 X_c \ket{\nb_{cb}}
  &=&
  \bra{- \nb_{bc}}e^{i\pi{\bf e}_z\cdot\sigma}(e^{-i\pi{\bf e}_z\cdot\sigma} X_b^\dagger)
 (X_c e^{i\pi{\bf e}_z\cdot\sigma})e^{-i\pi{\bf e}_z\cdot\sigma}\ket{\nb_{cb}} \nn\\
&=&
\bra{ \nb_{bc}} J^\dagger  e^{i\pi{\bf e}_z\cdot\sigma}\tilde X_b^\dagger
 \tilde X_c e^{-i\pi{\bf e}_z\cdot\sigma}g_{cb} J \ket{\nb_{bc}}\nn\\
&=&
\bra{ \nb_{bc}} J^\dagger e^{-i\phi} J^\dagger \tilde X_b^\dagger
 \tilde X_c g_{cb} J e^{i\phi} J \ket{\nb_{bc}} \nn\\
&=&
 \bra{ \nb_{bc}} J^\dagger J^\dagger \tilde X_b^\dagger
 \tilde X_c J \ket{
 \nb_{cb}}\nn\\
 &=& \overline{\bra{ -\nb_{bc}}\tilde X_b^\dagger
 \tilde X_c\ket{
 \nb_{cb}}}\eea
where we have defined the transformation $\tilde{X}_c = X_c e^{i\pi{\bf e}_z\cdot\sigma}$, which can be absorbed on
the group integration in \eqref{TheAmplitude} and the fact that $J\ket{
 \nb_{cb}}=\ket{
 -\nb_{cb}}$.
 We have used the Regge phase choice
\eqref{costateglue} from going from the first to the second line and the fact that the $\SU(2)$ transformations
are all in fact in the same $\U(1)$ subgroup (and hence commute).
From going to the second to the third line we have noted that we are acting with opposite rotations on the same state.
Hence we get that $\Z_{PR}(\Psi,\Sigma^3)= \overline{\Z_{PR}(\Psi, \Sigma^3)}$ which is thus real.

\section{Asymptotic formula}
%%%%%%%%%%%%%%%%%%%%%%%%
\label{asymptotic formula section}
We wish to study the semiclassical limit of the  amplitude $Z_{PR}(\Psi,\Sigma^3)$.  In order to do this, we
homogeneously rescale the spin labels by a factor $\lambda$.  The corresponding boundary state $\Psi_\lambda$ is
given by $\Psi_\lambda = \Psi( \lambda k_i , {\bf n_i}  )$.

Given a set $\mathcal{B}=\{ \mathbf n_{ab} , k_{ab} \}_{a \neq b}$ of  boundary data we denote as $\mathfrak{I}$
the set of cut immersions of the polyhedral surface $\partial \Sigma^3$ with edge lengths $k_{ab}$ in $\R^3$ up to
rigid motion.

A {\em cut immersion} $\mathfrak{i} \in \mathfrak{I}$ is an immersion of the manifold obtained from $\partial\Sigma^3$ by the trivializing cuts
$\partial\mathbb{D}_i$, $i\in C$, i.e. it is an immersion $\imath(\partial\Sigma^3-\{\cup_{i\in
C}\partial\mathbb{D}_i\})\hookrightarrow\R^3$. Furthermore, we require the existence of $\SO(3)$ elements that
identify the two sides of the cut, i.e.  $\hat{h}_i\in\SO(3)$ such that $\hat{h}_i(\imath(\partial\mathbb{D}_i^+))=
\imath(\partial\mathbb{D}_i^-)$, where $\partial\mathbb{D}_i^-$ and $\partial\mathbb{D}_i^+$ are the elements of
the boundary $\partial(\partial\Sigma^3-\partial\mathbb{D}_i)$ created by the removal of $\partial\mathbb{D}_i$
from $\partial\Sigma^3$  \footnote{Note that since $\mathbb{D}_i$ is transversal to generators of $H_1(\Sigma^3)$,
its removal changes the connectivity of $\Sigma^3$ and creates two new boundaries, $\mathbb{D}_i^-$ and
$\mathbb{D}_i^+$.} .

\begin{figure}
\begin{center}
\psfrag{h}{$\hat{h}\in \SO(3)$}
\includegraphics[scale=0.35]{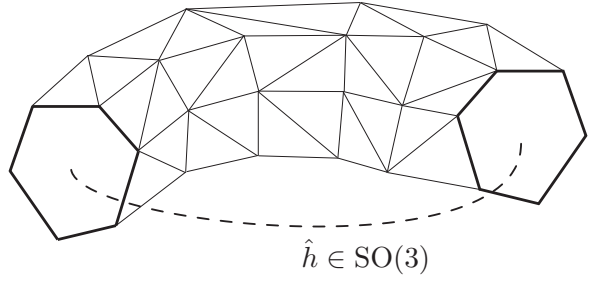}
\caption{A cut immersion for a particular boundary triangulation of a torus.   The cutting circles are shown in bold and there is an $\hat{h}\in \SO(3)$ that identifies them.}
\label{cut immersion}
\end{center}
\end{figure}

Any two cut immersions of $\Sigma^3$ are defined to be equivalent,  and can be  obtained from each other
if the cuts are related by a homotopy on the surface.
 Therefore, different choices of cuts
$\mathbb{D}_i$
%where $[\mathbb{D}_i]$ is dual to $[c_i]\in H_1(\Sigma)$
lead to equivalent cut immersions. An example of a cut immersion is given in Figure \ref{cut immersion}.
\begin{comment}
Such an immersion is called rigid if every continuous deformation of it  requires changing the edge lengths, and
flexible otherwise.
We denote the subset of  rigid immersions $\mathfrak{I}_r \subset \mathfrak{I}$.
%We denote the subset of flexible immersions $\mathfrak{I}_f \subset \mathfrak{I}$ and the subset $\mathfrak{I}_r \subset \mathfrak{I}$ and the .
Through every immersion in $\mathfrak{I}$ passes at least one manifold of flexible
immersions that can be continuously deformed into each other.
We call these { \em flexifolds} and denote them $\mathfrak{f}$, we denote the set of flexifolds $\mathfrak{F}$.
%We then define $\mathfrak{F}_{\mbox{\tiny max}}$ to be the set of maximally connected unions of flexifolds in $\mathfrak{F}$ containing at least one flexifold of maximal dimension.
We then define $\mathfrak{F}_{\mbox{\tiny max}}$ to be the set of flexifolds in $\mathfrak{F}$ of maximal dimension $d_{\mbox{\tiny max}}$.
With this definition the rigid immersions are a special case of flexible immersions with $d_{\mbox{\tiny max}}=0$.
We assume from now on that $\mathfrak{f}$ are manifolds.
\end{comment}

Such an immersion is called rigid if every continuous deformation of it  requires changing the edge lengths, and
flexible otherwise.
We denote the subset of  rigid immersions $\mathfrak{I}_r \subset \mathfrak{I}$.
%We denote the subset of flexible immersions $\mathfrak{I}_f \subset \mathfrak{I}$ and the subset $\mathfrak{I}_r \subset \mathfrak{I}$ and the .
Through every immersion in $\mathfrak{I}$ passes at least one manifold (with dimension $d$) of immersions that can be continuously deformed into each other.
We call these {\em flexifolds} and denote them $\mathfrak{f}$, we denote the set of flexifolds $\mathfrak{F}$.
%We then define $\mathfrak{F}_{\mbox{\tiny max}}$ to be the set of maximally connected unions of flexifolds in $\mathfrak{F}$ containing at least one flexifold of maximal dimension.
We then define $\mathfrak{F}_{\mbox{\tiny max}}$ to be the set of flexifolds in $\mathfrak{F}$ of maximal dimension $d_{\mbox{\tiny max}}$.
With this definition the rigid immersions are a special case of a flexifold with dimension $d=0$.
We assume from now on that the flexifolds $\mathfrak{f}$ do not intersect.

In the limit $\lambda \rightarrow \infty$ we have the following theorem:

\begin{theo} {\bf (Asymptotic formula)}
\label{big theorem new}
\begin{enumerate}
  \item If $\mathfrak{I}$ is not empty we have that:

\bea
\Z_{PR}(\psi_\lambda , \Sigma^3 )
&=&
\left(\frac{2 \pi}{\lambda}\right)^{\frac{3(|V |+|C| -1) - d_{\mbox{\tiny max}}  }{2}} \sum_{ \mathfrak{f} \in \mathfrak{F}_{\mbox{\tiny max}}} N_{\mathfrak{f}} \cos \left( \lambda
 \sum_{\substack{ ab \in E } } k_{ab} \Theta^{\mathfrak{f}}_{ab} + \phi_{ab}^\mathfrak{f} \right)
 \nn \\
 && +
 O\left(\left(\frac{1}{\lambda}\right)^{\frac{3(|V |+|C|) - d_{\mbox{\tiny max}}}{2} }\right)
\eea

%where $\mathfrak{I}_{f_{\mbox{\tiny max}}}\subset \mathfrak{I}_f$ consists of the elements in the highest
%dimensional strata of the flexifolds, with dimension $d_{\mbox{\tiny max}}$, and $N_{\mathfrak{i}}$ now also
%containing the volume of the flexifold.
The coefficient $N_{\mathfrak{f}}$, the dihedral angle $\Theta^{\mathfrak{f}}_{ab}$ and the phase $\phi_{ab}^\mathfrak{f}$ are independent of $\lambda$.
 %$\Theta^{\mathfrak{f}}_{ab}$ are  the dihedral angles of the cut immersion of the surface.
 The $\phi_{ab}^\mathfrak{f}$ and the dihedral angle $\Theta^{\mathfrak{f}}_{ab}$ are evaluated on an arbitrary immersion $\mathfrak{i}$ in $\mathfrak{f}$.
 It can be shown that these are independent of the cuts. Thus for any particular edge we can evaluate the dihedral angle by
 moving the cut away from it.
$|V |$ is the number of triangles (or equivalently vertices in the set $V$) and $|C|$ is the number of cutting circles.
$d_{\mbox{\tiny max}}$ is the dimension of the flexifolds $\mathfrak{f} \in \mathfrak{F}_{\mbox{\tiny max}}$, and $N_{\mathfrak{f}}$ now also
contains an integral over the union of flexifolds in $\mathfrak{f}$.

\item If no immersions in $\R^3$ exist the amplitude is exponentially suppressed:

\be
\Z_{PR}(\psi_\lambda , \Sigma^3 ) = o(\lambda^{-n}) \ \ \ \forall n
\ee

\end{enumerate}
\end{theo}

Note that in the simple case where the boundary data only admits rigid immersions, ie if $d_{\mbox{\tiny max}}=0$,
%if $\mathfrak{I}_f$ is empty and $\mathfrak{I}_r$ is not
 then the sum becomes a sum over the rigid immersions $\mathfrak{i} \in \mathfrak{I}_r$ and we have that:
\bea
\Z_{PR}(\psi_\lambda , \Sigma^3 )
&=&
\left(\frac{2 \pi}{\lambda}\right)^{\frac{3(|V |+|C| -1)}{2}}
 \sum_{ \mathfrak{i} \in \mathfrak{I}_r} N_{\mathfrak{i}} \cos  \left(  \lambda  \sum_{\substack{   ab \in E }  } k_{ab} \Theta^{\mathfrak{i}}_{ab} + \phi_{ab}^\mathfrak{i}       \right)
 \nn \\
 &&+ O\left(\left(\frac{1}{\lambda}\right)^{\frac{3(|V |+|C|- 1)}{2} + 1}\right)
\eea
since $d_{\mbox{\tiny max}} = 0$.
Since the immersions are now rigid, the coefficient $N_{\mathfrak{i}}$, the dihedral angles $\Theta^{\mathfrak{i}}_{ab}$ and the phase $\phi_{ab}^\mathfrak{i}$ are evaluated on the cut immersion $\mathfrak{i}$.

\section{Proof of the asymptotic formula}
%%%%%%%%%%%%%%%%%%%%%%%%
We now prove the above theorem.   We begin by describing the methods used to give the asymptotic form of the
amplitude, this will require finding the so-called stationary and critical points of the action.  We can then
interpret these points geometrically and give the asymptotic formula.  Much of this section is similar to \cite{Barrett:2009gg}
but one dimension lower so the analysis will be as brief as possible.

\subsection{Stationary phase}
%%%%%%%%%%%%%%%%%%%%%%%%
To find the asymptotic form of the  amplitude $I$, we will use the complex stationary phase formula
\cite{Hormander}.  The details of this are recalled below.

Let $D$ be a closed manifold of  dimension $n$, and let $S$ and $a$ be smooth,  complex valued functions on $D$
such that the real part $\Reel S \le 0$. Consider the function
 \be\label{integral}
 f(\lambda) = \int_D d x \, a(x) \, e^{\lambda
S(x)}. \ee The Hessian of $S$ is the $n\times n$  matrix of second derivatives of $S$ denoted $H$. For now let us
assume that the stationary points are isolated, which, by the Morse lemma is a condition equivalent to the
statement that the Hessian is  non-degenerate at the critical points; $\det H \neq 0$. Such functions are called
Morse functions.

In the extended stationary phase, the key role is played by {\em critical points},  that is, points $x_0$, which
are not only stationary: $\delta S(x_0)=0$ but for which $\Reel S(x_0)=0$ as well. So to compute the dominant terms in
the asymptotics for large spins, we need to find the stationary points of the action $S$ and restrict to those
with zero real part. If $S$ has no critical points then for large parameter $\lambda$ the function $f$ decreases
faster that any power of $\lambda^{-1}$. In other words, for all $N \geq 1$: \be \label{formula} f(\lambda) =
o(\lambda^{-N}), \ee

If there are isolated critical points, then each critical point contributes to the asymptotics of $f$ by a term of
order $\lambda^{-n/2}$. For large $\lambda$ the asymptotic expansion of the integral yields for each critical
point \be \label{formula1} a(x_0) \left(\frac{2 \pi}{\lambda}\right)^{n/2} \frac 1{\sqrt{ \det (-H)}} \,
e^{\lambda S(x_0)} \left[1+ O(1/\lambda) \right]. \ee
%where the index $\mathrm{Ind} \, H$ of $H$ is obtained as the sum of the argument of the eigenvalues $\rho_k$ of $H$, $\mathrm{Ind} \, H=\sum_{k=1}^n \mathrm{arg} \, \rho_k$, with $- \pi < \mathrm{arg} \rho_k \leq \pi$.

At a critical point, the matrix $-H$ has a positive definite real part,  and the square root of the determinant of
this matrix is the unique square root which is continuous on matrices with positive definite real part, and
positive on real ones. If $S$ admits several isolated critical points with non-degenerate Hessian, we obtain a sum
of contributions of the form \eqref{formula1} from each of them.

If there are also degenerate critical points, more care is needed to compute the asymptotics (see eg \cite{ramacher-2009}).
 Let $\mathcal{C}:=\{y\in D~|~ \delta S(y)=0,~\Reel S(y)=0 \}$ denote the set of critical points.
 %In our case, the action $S:\SU(2)^{|V|+|C|}\rightarrow \C$ is the function whose critical points we would like to study.
 %We will show that in the context of immersions above, $\mathcal{C}$ is related to $\mathfrak{I}_f$.
 %Note that we cannot a priori assume $\mathcal{C}$  to be itself the disjoint union of
%manifolds of the form $S^{-1}(S(y))$, since by definition $y$ are not a regular points of $S$.
For the action $S:\SU(2)^{|V|+|C|}\rightarrow \C$ we will show that $\mathcal{C}$ is the set of immersions $\mathfrak{I}$.
 Note that we cannot a priori assume $\mathcal{C}$  to be a disjoint union of manifolds as the flexifolds in $\mathfrak{I}$ can intersect. However, here we have restricted ourselves to this case so that the following generalized stationary phase theorem applies.

%There exists an extension of the above mentioned stationary formula for \eqref{integral} to the
%restricted case of

For a smooth function $S$ whose critical set $\mathcal{C}$ is a disjoint union of closed manifolds\footnote{In the literature this is called a
Morse-Bott function. A Morse function is the special case where the critical manifolds are zero-dimensional (so
the Hessian at critical points is non-degenerate in every direction, i.e., has no kernel).},
 each critical manifold $\mathcal{C}_{x_0}$ of dimension $p$, labelled by some $x_0$ on the critical manifold, contributes the following to the asymptotic formula \cite{ramacher-2009}:
\be
\label{formula2}
\left(\frac{2 \pi}{\lambda}\right)^{(n-p)/2}
e^{\lambda S(x_0)}
\int_{\mathcal{C}_{x_0}} d\sigma_{\mathcal{C}_{x_0}}(y) \frac {a(y)}{\sqrt{ \det (-H^\top(y))}} \,
\left[1+ O(1/\lambda) \right].
\ee
where $H^\top(y)$ is the restriction of the matrix to the directions normal to
$\mathcal{C}_{x_0}$ with respect to some Riemannian metric on the domain
%(which in our case is the configuration space $\SU(2)^{|V|+|C|}$)
, and $d\sigma_{\mathcal{C}_{x_0}}$ is the canonical measure induced on the critical submanifold by the
same Riemannian measure on the domain space.
This extends to the case where $\mathcal{C}$ is a manifold-with-boundary.
 %We have assumed that our boundary data admits at most such types of degenerate critical points.

\subsubsection{Critical points}
%%%%%%%%%%%%%%%%%%%%%%%%
\label{critical points section}

As described above, we must find the  points of the action \eqref{action} such that $\Reel S = 0$ as these are the
only points that contribute in the limit $\lambda \rightarrow \infty$.  First, we introduce some more notation.
The action of the elements $X_b$ on the coherent states will produce a new coherent state \be
  |  {\bf n}_{ab}'   \rangle = X_{a}   |  {\bf n}_{ab}   \rangle
\ee We will denote the corresponding rotated  three vectors by \be
  {\bf n}_{ab}'  = \hat{X}_{a}   { \bf n}_{ab}
\ee
where $\hat{X}_{a}$ is the $\SO(3)$ element corresponding to $X_a$.

We will first consider critical  points for edges that are not on one of the cutting circles.  Using \eqref{coh
state mod sq}, we can see that the real part of the action is given by
\be
\Reel S = \sum_{ab \in \tilde{E}}
k_{ab} \ln \frac{1}{2}(1 - {\bf n}_{ab}' \cdot {\bf n}_{ba}'  ).
\ee
This does not depend on the coherent state
phases as it is real.  Using this formula, we can see that $\Reel S = 0$ when ${\bf n}_{ab}' =   - {\bf n}_{ba}'$
for all $ab$, or explicitly in terms of $\SO(3)$ rotations
\be
\label{critical points}
 \hat{X}_{a}
 { \bf  n}_{ab}   =  - \hat{X}_{b}    { \bf n}_{ba}  .
\ee The critical points for an edge that crosses a cutting circle $i$ differ by the inclusion of the $h_i$ \be
\label{critical points on C} \hat{X}_{a}    { \bf n}_{ab}   =  - \hat{h}_i \hat{X}_b    { \bf n}_{ba}  . \ee

\subsubsection{Stationary points}
%%%%%%%%%%%%%%%%%%%%%%%%
The stationary points are found by varying  the action with respect to each of the group variables $X_a$.  The
variation of an $\SU(2)$ group variable and its inverse is
\be \delta X = T X \ \ \ \ \ \,       \delta
X^{-1} = - X^{-1}  T
\ee for an arbitrary $\mathfrak{su}(2)$ Lie algebra element $T = \frac{1}{2} i T^j
\sigma_j$. The stationary points are given by $\delta S = 0$ and lead to the following equation
\be \sum_{b\colon
b \ne a} k_{ab} \,\, \mathbf{V}_{ab} = 0
\ee
where
\be
\label{vector}
\mathbf{V}_{ab}
= \frac{\la -\mathbf n_{ab}
|X_a^{-1} \, {\pmb\sigma} \, X_b |   \mathbf n_{ba} \ra}{\la -\mathbf n_{ab} | X_a^{-1} X_b |   \mathbf n_{ba}
\ra},
\ee
These equations can then be evaluated at the critical points to give
\be \sum_{b\colon b \ne a} k_{ab}
\,\,{\nb_{ab}} = 0
\ee
which is the closure constraint for an immersed triangle.

The stationary phase condition for the $h_i$ variables is the same but in this case we obtain
\be \label{circleclosure}
\sum_{ab \in
C_i} k_{ab} \,\,{\bf n}_{ab}' = 0
\ee
Which is the closure condition for edges on the circle $i$ immersed in
$\mathbb{R}^3$. Note that unlike the closure condition for the triangle, this relation involves the ${\bf n}_{ab}'$ as each edge belongs to a different triangle.

If the critical points are not isolated but form a manifold of critical points, then we denote this manifold by
\be
\mathcal{C}_{X} = \{ (X_1 , ... , X_{|V|},h_1, ... ,h_{|C|})      \in \SU(2)^{|V|+|C|}  :   \delta S=0 , \Reel (S) =0            \}
\ee

\subsection{Geometry}
We will now describe how the  critical/stationary points can be given a geometric interpretation. We first consider the case where the critical points are isolated.
\begin{theo}{\bf (Geometry)}
\label{reconstruction}
Given a set of boundary data  $\mathcal{B}$ satisfying the closure constraint on each
triangle, the solutions $X_a, h_i$ to the critical and stationary point equations \eqref{critical points}, \eqref{critical points on C} and \eqref{circleclosure} correspond to the oriented
geometric immersions $\imm$ of a geometric triangulated 2-manifold with boundary $\bigcup_{i\in C}
\partial\mathbb{D}^+_i \cup \partial\mathbb{D}^-_i$ obtained by a suitable cutting of the boundary manifold.
This immersion is subject to the constraint that a set of $\hat{h}_i \in \SO(3)$ exist that map the immersion of
$\partial\mathbb{D}^+_i$ to the parity flipped $P (\partial\mathbb{D}^-_i)$, that is, the immersion of
$\partial\mathbb{D}^+_i$ is congruent and oppositely oriented to the immersion of $\partial\mathbb{D}^-_i$.

The geometric vectors of the  immersion are given by
$${\bf v}_{ab} (\imm) =  k_{ab} \hat{X}_a \nb_{ab}.$$
 Its orientation the one induced by the vectors on each face.

Conversely, an immersed surface $\imm$ determines a set of $k_{ab}, \nb_{ab}$ and a set of $\SO(3)$ elements $\hat{X}_a(\imm)$.
\end{theo}

Proof: Start somewhere on  the surface of $\Sigma^3$ that is not in a circle. Since $\Sigma^3$ has connected
boundary, the entire surface is now contractible to this point if cut along the circles. Take the triangle
$\Delta_a$ you are on as embedded in $\pole^\bot$ and rotate it according to the stationary point equations to
$X_a \Delta_a$. Embed the next triangle and rotate it, according to the stationary point equations it's edges are
now antiparallel to already immersed edges. As they are geometrically glued a translation exists that identifies
all its edges with already immersed ones. Thus iteratively the whole immersed surface can be built up and the closure conditions on the cuts now imply that the surface closes up. Finally the stationarity equations on the circles imply that the $P \hat{h}_i$ identifies the circles where we cut the surface.

Conversely given an oriented  immersion with the right edge lengths we can choose a set of edge vectors compatible
with the orientation on the surface. On each triangle there are two linearly independent edge vectors. The map
from these to the corresponding boundary elements defines a rotation in $\SO(3)$. On the boundary circles we
explicitly have $\SO(3)$ elements. The complete set of these solves the critical point equations. $\square$

If there is a manifold of dimension $d>0$ of critical points then Theorem \ref{reconstruction} holds for each critical point in $\mathcal{C}_X$.  Since these critical points lie on a manifold, there is a continuous deformation of the immersed surface that does not change the edge lengths.  Hence these critical points reconstruct flexible immersions and we arrive at the flexifolds $\mathfrak{f}$ described in section \ref{asymptotic formula section}.  We will now label the critical manifolds by $\mathcal{C}_\mathfrak{f}$, where $\mathfrak{f}$ is the flexifold that it describes.

\paragraph*{Geometrical interpretation}
We will now describe in some more detail  the geometric structure of these surfaces. Given an oriented surface in
$\R^3$ the standard orientation automatically gives us a consistent set of normals $\pole_a$. By our choice of boundary data
we have automatically ensured that these are given simply by
$$\pole_a = \hat{X}_a \pole.$$
We can define the dihedral rotation for an oriented surface unambiguously as the rotation $\hat{D}_{ab} \in \SO(3)$ around the geometric edge vector
$\vb_{ab}(\imm)$ that takes $\pole_a$ to $\pole_b$, that is
$$\pole_b = \hat{D}_{ab} \pole_a$$
and
$$\vb_{ab}(\imm) = \hat{D}_{ab} \vb_{ab}(\imm).$$
the lift of this rotation can thus be written as
\bea
\label{dihedral as exp}
D_{ab} &=&  \exp{  \left( \Theta_{ab}^\imm \, \frac{ \vb_{ab}(\imm) }{|\vb_{ab}(\imm)|}    .L  \right)}
\nn \\
&=& \exp{(      \Theta_{ab}^\imm \, \nb_{ab}'.L)}
\eea
where we require
$-\pi < \Theta_{ab}^\imm \leq \pi$. We then call $\Theta_{ab}^\imm$ the dihedral angle. As $\vb_{ab}(\imm) = - \vb_{ba}(\imm)$
this definition clearly implies $\Theta_{ab}^\imm = \Theta_{ba}^\imm$. If we have a surface defining a convex subspace of
$\R^3$ this definition reduces to the usual definition. In particular the consistent choice of orientation then
ensures that we have $0 < \Theta_{ab}^\imm \leq \pi$ and $\cos{(\Theta_{ab}^\imm)} = \pole_a\,.\,\pole_b$ for outward facing
normals.

On the boundaries of the surface we can  also define an analogue of the dihedral rotation by requiring $$\hat{h_i}
\pole_b = \hat{D}_{ab} \pole_a$$ and $$ - \hat{h}_i \vb_{ba}(\imm)= \vb_{ab}(\imm) =\hat{D}_{ab}
\vb_{ab}(\imm)$$ Geometrically this makes sense as it corresponds to the angle obtained by gluing on the two
identified boundaries.

\paragraph*{Commuting diagram}

To fully connect the geometry of these  surfaces to the boundary we can relate the dihedral rotation to the gluing
of the boundary $g_{ab}$ defined above.

\label{commuting diagrams} Consider the following diagram which applies to two adjacent triangles that are not on
a cutting circle:
\begin{equation}
\label{commutative}
\xymatrix{\ar @{} [dr]  t_a \ar[d]_{g_{ab}} \ar[rr]^{X_a} && ~\tau_a \ar[d]^{(-1)^{\nu_{ab}} D_{ab}}   \\
t_b\ar[rr]_{X_b} && ~\tau_b  }
\end{equation}

Here $t_a$ is  the boundary triangle at $\pole^\bot$ with edge vectors given by $ k_{ab} \nb_{ab}$ and $\tau_a$ is the
triangle rotated according to its location in the surface, which according to the reconstruction theorem
\ref{reconstruction} has edge vectors given by $\vb_{ab}(\imm) =  k_{ab} \hat{X}_a \nb_{ab}$. The $\SO(3)$ action of
this diagram immediately commutes, as can be seen by acting on $\nb_{ab}$ and $\pole$. As an $\SU(2)$ diagram this
equation defines a sign $(-1)^{\nu_{ab}}$ that makes it commute. The discrete sign symmetry
$X_a\rightarrow\epsilon_aX_a$ of the action can be seen as acting on this sign by $(-1)^{\nu_{ab}} \rightarrow
\epsilon_a \epsilon_b (-1)^{\nu_{ab}}$.

Now, for two triangles whose common edge is on a cutting circle $i$,  in the same way we have a commuting diagram
as
\begin{equation}
\label{commutative with cut}
\xymatrix{\ar @{} [dr]  t_a \ar[d]_{g_{ab}} \ar[rr]^{X_a} && ~\tau_a \ar[d]^{(-1)^{\nu_{ab}} D_{ab}}   \\
t_b\ar[rr]_{h_i X_b} && ~\tau_b  }
\end{equation}
and additionally have $(-1)^{\nu_{ab}} \rightarrow \epsilon_a \epsilon_b \epsilon_i (-1)^{\nu_{ab}}$.

We now show that the dihedral rotation is unchanged by moving the cut.
Suppose that $ab\in i\in C$, in the sense that the edge $\vb_{ab}(\imm)$ crosses the cutting circle labelled
$i$. Now choose a different, yet homotopic, cutting circle $\partial\widetilde{\mathbb{D}}_i\sim
\partial\mathbb{D}_i$, such that now $ba\notin i$, but $cb\in i$. This corresponds to sliding the cut from one
edge of $\tau_b$ to another, and as any other change in the cut,  it will modify wether labels $a$ and $b$ satisfy
\eqref{critical points} or \eqref{critical points on C}. By our orientation convention, we now have
\be
{X}_{a} { \bf n}_{ab}   =  -\widetilde{X}_b    { \bf n}_{ba}~~~~\mbox{,}~~~~  - \widetilde{h_i}
\widetilde{X_c}\nb_{cb} = {X_b}\nb_{bc} \ee By just looking at the first set of equations above,
and comparing it to ${X}_{a} { \bf n}_{ab} = -h_i{X}_b { \bf n}_{ba}$ we have that $h_i{X}_b    { \bf n}_{ba}=
\widetilde{X}_b { \bf n}_{ba}$. This shows that $h_i{X}_b$ and $\widetilde{X}_b$ are equal up to a phase,
 and by the reconstruction theorem in fact $h_i{X}_b=\widetilde{X}_b$. Hence,
comparing the diagrams \eqref{commutative} and \eqref{commutative with cut} we get that moving the cut does not
affect the dihedral angle as here defined.

\paragraph*{Parity}

Given an oriented surface immersed in $\R^3$  all surfaces related to it by $\OO(3)$ clearly have the same
boundary data. Those related by $\SO(3)$ also have the same dihedral angle as defined above. However if we act by
parity $P: \nb \rightarrow -\nb$ we switch the dihedral angle. This is because the two equations defining $D_{ab}$
are invariant under parity, and the dihedral rotation is unchanged. Thus by the definition of the dihedral angle
we have
$$
D_{ab} = \exp{\left( \Theta_{ab}^\imm \, (- \frac{\vb_{ab}}{|\vb_{ab}|}   (P\sigma)).L  \right)}
$$
and so  $\Theta_{ab}^{P \imm}  = - \Theta_{ab}^\imm $.

\subsection{The Regge action}

Now note that given  a geometric surface and associated solution $X_a(\imm)$, we can obtain the solution $X_a(P
\imm)$ corresponding to the parity flipped surface explicitly. That is, by equation \eqref{dihedral as exp} we
have that \be \label{parity} (X_a(\imm))^{-1} D_{ab} X_a(\imm) = \exp{( \Theta_{ab}^\imm \, \nb_{ab}.L)} \ee
Acting with $X_a^{-1}$ by the left of the commuting diagram equations, using the notation $X_{ab}(\imm) =
(X_a(\imm))^{-1} X_b(\imm)$ if not on a circle and $X_{ab}(\imm) = (X_a(\imm))^{-1} h_i X_b(\imm)$ if on, we get
\be \label{gab} X_{ab}(\imm) g_{ab} = (-1)^{\nu_{ab}}(X_a(\imm))^{-1} D_{ab} X_a(\imm) =(-1)^{\nu_{ab}}
\exp{(\Theta_{ab} \, (\nb_{ab}.L)}. \ee where we have used \eqref{parity}.

It is now straightforward to evaluate  the matrix elements in the amplitude. These are of the form
$\bra{-\nb_{ab}} X_{ab} \ket{\nb_{ba}}$. Using the gluing condition this becomes $\bra{-\nb_{ab}} X_{ab} g_{ab}
\ket{- \nb_{ab}}$. Finally, by \eqref{gab}
 this is just $(-1)^{\nu_{ab}} e^{\frac{i}{2}
\Theta_{ab}^\imm }$. Thus we have overall that: \be \label{action with lift} \bra{-\nb_{ab}} X_{ab} \ket{\nb_{ba}}
= (-1)^{\nu_{ab}} e^{\frac{i}{2} \Theta_{ab}^\imm} \ee

\paragraph{Spin Structure}

Now we will fix the ambiguity in signs emerging from the spin lifts of the dihedral angle by exploring the
discrete sign  symmetry in $h_i$ and $X_a$. Recall that the discrete sign freedom of the action
$X_a\rightarrow\epsilon_aX_a$ emerged from a different choice of spin frame for each triangle. Now, we show that
the discrete sign symmetry related to the cuts $h_i\rightarrow\epsilon_ih_i$ corresponds to different choices of
spin structures for the manifold $\Sigma^3$. Then, using the fact that $(-1)^{\nu{ab}} D_{ab}$ is a gauge
transform of the connection $g_{ab}$ we can fix the symmetries by adjusting the spin frames and the spin structure
such that $(-1)^{\nu_{ab}} = 1$. Thus we will show that:
\begin{lem}
\label{spin lift lemma} The signs $(-1)^{\nu_{ab}}$ arising from the  spin lift on each face not on the cut obey
$(-1)^{\nu_{ab}}=\kappa_{ab}=\kappa_a\kappa_b$ for some $\kappa_a=\pm 1$. The signs for a face on the cut, i.e.
$ab\in i\in C$ obey $(-1)^{\nu_{ab}}=\kappa_{ab}=\kappa_a\kappa_b\kappa_i$ where $\kappa_i$ parametrizes the spin
structures of $\Sigma^3$.
\end{lem}

{\em Proof.} First of all,  by \eqref{commuting diagrams} a lift of the dihedral rotations, $\kappa_{ab}{D}_{ab}$,
are just a gauge transformation of the ${g}_{ab}$. Now recall that $\hat{g}_{ab}\in\SO (3)$ are parallel
translations on the boundary triangles according to the Levi-Civita connection of the associated metric, with
 $g_{ab}$ being the parallel translation of the respective spin connection (a lift of $\hat{g}_{ab}$ to $\SU (2)$).

 But when the geometry
 around a vertex is continuously deformed to the flat geometry, the $g_{ab}$ holonomy of a trivial cycle around said
 vertex has to go to the identity rotation, as opposed to a $2\pi$ rotation.
  This implies that for the holonomy around a vertex through triangles $a,b$ and $c$
  (which of course consists of a trivial cycle), we have
 $$\kappa_{ca}\kappa_{bc}\kappa_{ab}{D}_{ca}{D}_{bc}{D}_{ab}=\kappa_{ca}\kappa_{bc}\kappa_{ab}=1 $$
which implies that locally we must have $\kappa_{ab}=\kappa_a\kappa_b$. The problem now is that if there are
non-trivial cycles, i.e. $g\neq0$, we may not be able to extend this globally, i.e. $\kappa_{ab}$ may not be
globally pure gauge.

 In other words, for trivial cycles
  the lift of the holonomy given by the $\kappa_{ab}D_{ab}$  is fixed
   to be the same as that given by $D_{ab}$.
  But not so for the holonomy of a non-trivial cycle; there exist inequivalent spin structures
 on a manifold. These have a one-to-one correspondence with the elements of
$H_1(\Sigma^3,\mathbb{Z}_2)$, and so are $2^g$ in number. Hence for a non-trivial cycle, dual to
 the sequence of triangles $\Delta_{a_0}\cdots \Delta_{a_n}\Delta_{a_0}$ crossing the circle $i\in C$, we have
\be\label{Dihedral
chain}\kappa_{a_na_0}\kappa_{a_{n-1}a_{n}}\cdots\kappa_{a_0a_1}{D}_{a_na_0}{D}_{a_{n-1}a_{n}}\cdots{D}_{a_0a_1}
  =\kappa_i{D}_{a_na_0}{D}_{a_{n-1}a_{n}}\cdots{D}_{a_0a_1}\ee where a $\kappa_i$ is introduced whenever there
  is an implicit choice of spin structure; i.e. it parametrizes the different
  spin structures associated with the cut.

We reconcile this case with the $g=0$ one by keeping the form $\kappa_{ab}=\kappa_a\kappa_b$ for all the edges
$ab$ that do not lie on a circle, i.e. $ab\notin i$ for any $i\in C$. Then by \eqref{Dihedral chain} immediately
we must have for $ab\in i$, $\kappa_{ab}=\kappa_i\kappa_a\kappa_b$. Since our chosen basis for
$H_1(\Sigma^3,\mathbb{Z}_2)$ generates all cycles, we can see that this form of $\kappa_{ab}$ has all the right
properties demanded by our equations and accounts for the different spin structures.

Therefore, taking advantage of the discrete sign symmetry, we can choose the spin structure to be compatible with
the one chosen for the lift of $g_{ab}$ and thus we will have $(-1)^{\nu_{ab}} \rightarrow \epsilon_i\epsilon_a
\epsilon_b (-1)^{\nu_{ab}}$ makes $(-1)^{\nu_{ab}}=1$.
 $\square$.

Finally, we obtain that the action evaluated at the critical points is the Regge action for the immersed surface
$\mathfrak{i}$. \be S = \sum_{ab \in E} k_{ab} \Theta^\mathfrak{i}_{ab} \ee For the flexible immersions, the
action is the same for all points\footnote{This is actually a particular case of the ``strong bellows conjecture''
that was shown in \cite{}.} on the critical manifold so we evaluate it on an arbitrary immersion in the flexifold.

This concludes the derivation of the Regge action.

\subsection{Hessian}

The stationary phase formula requires us to calculate the Hessian  of the action $S$ to determine the weights with
which the stationary points contribute to the action. This  will be a $3(|V| +|C|) \times 3(|V|+|C|)$ matrix defined
by
\be
H
=
\left(
\begin{array}{cc}
  H_{XX} & H_{Xh} \\
  H_{hX} & H_{hh}
\end{array}
\right).
\ee
Where
\bea
(H_{XX})^{ij}_{cd} = \left(\frac{\partial^2S}{\partial X^{i }_c\partial X^{j}_d}\right)  \\
(H_{hX})^{ij}_{pd} = \left(\frac{\partial^2S}{\partial h^{i }_p\partial X^{j}_d}\right) \\
(H_{Xh})^{ij}_{cq} = \left(\frac{\partial^2S}{\partial X^{i }_c\partial h^{j}_q}\right)  \\
(H_{hh})^{ij}_{pq} = \left(\frac{\partial^2S}{\partial h^{i }_p\partial h^{j}_q}\right)
\eea
The global $\SU(2)$ symmetry of the action implies that there is a redundant integration in $I$.  This will cause the determinant of
the Hessian to be zero unless it is gauge fixed.  To solve this, we make the change of variables $X_a \rightarrow
X_b X_a$ for some $b \in \{ 1,..., |V| \}$. This has the effect of removing the $X_b$ variables and the integral gives a volume of
$\SU(2)$ which can be normalised to one as it is compact. The remaining Hessian is now a $3(|V|+|C|-1) \times
3(|V|+|C|-1)$ matrix.
The submatrix $H_{XX}$ at the critical points, is given by
 \be
\left.\left(\frac{\partial^2S}{\partial X^{i}_c\partial X^{j}_c}\right)\right|_{\delta S = 0 ,\Reel S=0}=
\frac{1}{2}\sum_{b  \neq c, bc \in E}k_{cb}\left(\delta^{ij}-{n'}^{i}_{cb}{n'}^{j}_{cb}\right) \ee
for the diagonal terms.
The off diagonal part is
\be \left. \left(\frac{\partial^2S}{\partial X^{i}_c\partial
X^{j}_d}\right)\right|_{ \substack{   \delta S = 0 \\ \Reel S=0}}
= - \frac{1}{2} \sum_e   ( \delta_{c~s(e)}\delta_{d~t(e)}  + \delta_{d~s(e)}\delta_{c~t(e)}  )
\left(\delta^{ij}-i\epsilon^{ijk} n^{k}_{s(e)t(e)} -{n'}^{i}_{s(e)t(e)}{n'}^{j}_{s(e)t(e)}\right).
\ee
So one can see that only the off-diagonal elements that represent two neighbouring triangles are non zero.
The $(H_{hh})$ submatrix will be diagonal since each term in the action only contains one $h_p$ term (ie, each dual edge only crosses one cut.)
\be
 \left.  \left( \frac{\partial^2S}{\partial h^{i }_p \partial h^{j}_p} \right)      \right|_{\delta S = 0 ,\Reel S=0}
=
\frac{1}{2}\sum_{b  \neq c, bc \in C_p} k_{cb} \left( \delta^{ij}-{n'}^{i}_{cb}{n'}^{j}_{cb}\right)
\ee
The mixed terms $H_{Xh},H_{Xh}$ will be non zero only for triangles with an edge on the cut
\be
 \left. \left(\frac{\partial^2S}{\partial X^{i }_c\partial h^{j}_q}\right) \right|_{ \substack{   \delta S = 0 \\ \Reel S=0}}
=
- \frac{1}{2}
\sum_{  \substack{ab \in E_q \\ c=a,b}   }
k_{ab}\left(\delta^{ij}-i\epsilon^{ijk} n^{k}_{ab}
-{n'}^{i}_{ab}{n'}^{j}_{ab}\right)
\ee

Note that
\be
(X_a(\imm))^{-1} D_{ab} X_a(\imm)
= \exp{(-\Theta_{ab}^\imm \, (\nb_{ab}(\imm)).L)}^{-1}
= \left((X_a(P \imm))^{-1} D_{ab} X_a(P \imm)\right)^{-1}
\ee
where we used that
$D_{ab} = \exp{(\Theta_{ab}^\imm \,
(- \frac{\vb_{ab}(P\imm)}{|\vb_{ab}(P\imm)|} ).L)}$
on the second equality. By \eqref{gab} we then have $(X_{ab}(\imm) g_{ab})=(X_{ab}(P
\imm) g_{ab})^{-1}$, so if we replace the $X_{ab}(\imm)$ in $\bra{-\nb_{ab}} X_{ab}(\imm) \ket{\nb_{ba}}$
with the parity related one we now obtain the complex conjugate: \bea \bra{-\nb_{ab}} X_{ab}(\imm) g_{ab}
\ket{-\nb_{ab}} &=&
\bra{-\nb_{ab}} (X_{ab}(P \imm) g_{ab})^{-1} \ket{-\nb_{ab}} \nn \\
&=&
 \overline{\bra{-\nb_{ab}}  X_{ab}(P \imm) g_{ab}\ket{-\nb_{ab}}}
 \nn\\
 &=&
 \overline{\bra{-\nb_{ab}} X_{ab}(P \imm) \ket{\nb_{ba}}}\eea
Thus we can see that the action
of parity on the Hessian matrix will also result in complex conjugation when evaluated at the critical points.

\subsection{Proof of the formula}

We can now evaluate the stationary phase approximation to the amplitude $\Z_{PR}(\Psi_\lambda, \Sigma^3)$ defined in \eqref{TheAmplitude}. We begin by fixing the symmetries of the action. This can be achieved by taking an arbitrary vertex and dropping the group integration associated to it. As shown in section \ref{critical points section} the critical point equations are the equations for the immersion of a polyhedral surface with the geometry specified in the boundary data. If a particular immersion is rigid no infinitesimal deformation taking it to another such immersion exists and therefore it is an isolated critical point of the amplitude.

For the isolated critical points in $\mathfrak{I}_r$ we can explicitly evaluate the stationary phase approximation. Having fixed one group integration we are left with a $3(|V|+|C|-1)$ dimensional integration. The overall scaling of these points is thus $\left(\frac{2\pi}{\lambda}\right)^{3(|V|+|C|-1)/2}$. Further we obtain a set of $2^{3(|V|+|C|-1)}$ critical points for each immersion from the spin lift of each $\SU(2)$. Finally the derivatives in the Hessian as defined above are taken with respect to a parametrization of $\SU(2)$ with volume $(4\pi)^2$, so we need to rescale by this factor. Using equation \eqref{action with lift} and lemma \ref{spin lift lemma} the amplitude itself evaluates to the Regge action of the cut immersion:
$$
\ln \bra{-\nb_{ab}} X_{ab} \ket{\nb_{ba}}^{2k_{ab}} = i k_{ab} \Theta_{ab}^\imm.
$$
Taking all these factors together and using the fact that we know parity related immersions to be the complex conjugate of each other we can approximate the contributions of the isolated critical points to the partition function as:
\bea
\Z_{PR}(\Psi_\lambda, \Sigma^3)&=& (-1)^\chi
\left(\frac{2 \pi}{\lambda}\right)^{\frac{3(|V |+|C| -1)}{2}}
\left(\frac{2}{(4 \pi)^2}\right)^{\frac{3(|V |+|C| -1)}{2}}
 \sum_{ \mathfrak{i} \in \mathfrak{I}_r} \frac{1}{\sqrt{\det H_{\mathfrak{i}}}} \exp  \left( i  \lambda  \sum_{\substack{   ab \in E }  } k_{ab} \Theta^{\mathfrak{i}}_{ab}\right) \nn \\
 &+& \frac{1}{\sqrt{\overline{\det H_{\mathfrak{i}}}}} \exp  \left( - i  \lambda  \sum_{\substack{   ab \in E }  } k_{ab} \Theta^{\mathfrak{i}}_{ab}\right) + O\left(\left(\frac{1}{\lambda}\right)^{\frac{3(|V |+|C|- 1)}{2} + 1}\right)
\eea
where $\Theta^{\mathfrak{i}}_{ab}$ is the dihedral angle of the edge $ab$ in the cut immersion $\mathfrak{i} \in \mathfrak{I}_r$.
Since the Hessian matrix changes to its complex conjugate with parity, we can absorb the phase of the determinant into the exponentials and combine the terms
\bea
\Z_{PR}(\Psi_\lambda, \Sigma^3)&=& 2(-1)^\chi
\left(\frac{1}{4 \pi \lambda}\right)^{\frac{3(|V |+|C| -1)}{2}}
 \sum_{ \mathfrak{i} \in \mathfrak{I}_r}
 \frac{1}{\sqrt{|\det H_{\mathfrak{i}}|}}
 \cos \left( i  \lambda  \sum_{\substack{   ab \in E }  } k_{ab} \Theta^{\mathfrak{i}}_{ab}
 -\frac{1}{2}\mathrm{Arg}(\det H_{\mathfrak{i}})   \right)\nn \\
 & & + O\left(\left(\frac{1}{\lambda}\right)^{\frac{3(|V |+|C|- 1)}{2} + 1}\right).
\eea

If there are any flexible immersions of the boundary data then there will be a manifold of critical points.  Since the critical points extremize the action, it must have the same value on every point of the critical manifold.
The Hessian therefore has zero modes along the
directions of the flexifold and we must treat the integral as having further symmetries in the neighbourhood of
the flexifold.
Factoring out these changes the scaling of the contribution of these critical points by
$\lambda^{d_{\mbox{\tiny max}}/2}$, where $d_{\mbox{\tiny max}}$ is the dimension of the flexifold.
Therefore these immersions dominate the rigid immersions if they exist, their contribution is given by:
\bea
\Z_{PR}(\Psi_\lambda, \Sigma^3)
&=& (-1)^\chi
\left(\frac{2 \pi}{\lambda}\right)^{\frac{3(|V |+|C| -1) - d_{\mbox{\tiny max}}}{2}} \left(\frac{2}{(4 \pi)^2}\right)^{\frac{3(|V
|+|C| -1) - d_{\mbox{\tiny max}}}{2}}
\nn \\
&& \times
\left[
 \sum_{ \mathfrak{f} \in \mathfrak{F}_{\mbox{\tiny max}}} L^\mathfrak{f}
 \exp  \left( i  \lambda  \sum_{\substack{   ab \in E }  } k_{ab} \Theta^{\mathfrak{f}}_{ab}\right)
 + \overline{L^\mathfrak{f}} \exp  \left( - i  \lambda  \sum_{\substack{   ab \in E }  } k_{ab} \Theta^{\mathfrak{f}}_{ab}\right)
\right]
\nn \\
 &&+ O\left(\left(\frac{1}{\lambda}\right)^{\frac{3(|V |+|C|- 1) - d_{\mbox{\tiny max}}}{2} + 1}\right)
\eea
$\Theta^{\mathfrak{f}}_{ab}$ is the dihedral angle of the edge $ab$ of a particular cut immersion $\mathfrak{i}$ in the flexifold  $\mathfrak{f}$.
As the action is constant along the flexifold it does not matter where we evaluate it.
$L^\mathfrak{f}$ is given by
\be
L^\mathfrak{f} = \int_{\mathcal{C}_{\mathfrak{f}}} d\sigma_{\mathcal{C}_{\mathfrak{f}}}(y) \frac {a(y)}{ \sqrt{\det H^\top_{ \mathfrak{f} }(y)  }   }
\ee
where $H^\top_{ \mathfrak{f} }$ is the Hessian matrix for the transverse directions which we can not give a general formula for.
Combining the exponentials into cosines as above we obtain part one of the main theorem.

Finally, if no immersions of the boundary data exist then there are no solutions to the critical point equations and the stationary phase formula gives that the amplitude is suppressed. $\square$

\section{Example: The Tetrahedron}
\label{tet section}

Here we apply the above results to the well known case of the asymptotics of the amplitude for a single tetrahedron which, with an appropriate choice of normalisation for the boundary intertwiners, will correspond to the 6j symbol.
This is a special case of theorem \ref{big theorem new} so the proof is the same as above.  In particular, the critical and stationary point equations are the same and the action evaluated at these points reduces to the Regge action for a tetrahedron.
Since the asymptotic formula for the tetrahedron is already known, we must verify that our formula agrees with this result.  This also provides further evidence that the asymptotic formula for the 4d case derived using the same methods in \cite{Barrett:2009gg} is correct.
We begin by noting that, up to parity,  the boundary data of a tetrahedron has only one immersion so the sum in the asymptotic formula disappears.

We will also require an explicit formula for the Hessian matrix in order to perform the numerical calculations. For the tetrahedron, the Hessian is the $12 \times 12$ matrix defined by
\be
H_{cd}^{ij} = \left(\frac{\partial^2S}{\partial X^{i
}_c\partial X^{j}_d}\right).
\ee
The global $\SU(2)$ symmetry of the action implies that there is a redundant integration in $I$.  This will cause the determinant of the Hessian to be zero unless it is gauge fixed.  To solve this, we make the change of variables $X_a \rightarrow X_4 X_a$ for $a=1,2,3$.  This has the effect of removing the $X_4$ variables and the integral gives a volume of $\SU(2)$ which can be normalised to one as it is compact. The remaining Hessian is now a $9 \times 9$ matrix which, at the critical points, is given by
 \be
\left.\left(\frac{\partial^2S}{\partial X^{i}_c\partial X^{j}_c}\right)\right|_{\delta S = 0 ,\Reel S=0}=
\frac{1}{2}\sum_{b\neq c}k_{cb}\left(\delta^{ij}-{n'}^{i}_{cb}{n'}^{j}_{cb}\right)
\ee
for the diagonal terms.  The off diagonal part is
\be
\left.
\left(\frac{\partial^2S}{\partial X^{i}_c\partial
X^{j}_d}\right)\right|_{\delta S = 0 ,\Reel S=0}
=-\frac{1}{2}k_{cd}\left(\delta^{ij}-i\epsilon^{ijk} n^{k}_{cd}
-{n'}^{i}_{cd}{n'}^{j}_{cd}\right). \ee

\subsection{Normalisation and scaling behaviour}

We can now compare our theorem with the Ponzano-Regge asymptotic formula for the 6j symbol.  The Ponzano-Regge formula is
\be
\label{pr formula}
\left\lbrace
\begin{array}{ccc}
\lambda k_{12} & \lambda k_{13} & \lambda k_{14} \\
\lambda k_{23} & \lambda k_{24} &  \lambda k_{34}
\end{array}
\right\rbrace
\rightarrow
\frac{1}{\sqrt{12 \pi \mathrm{Vol} }} \cos \left(  \sum_{a<b} (\lambda k_{ab} + \frac{1}{2}) \Theta_{ab}    + \frac{\pi}{4}   \right)
\ee
where $\mathrm{ Vol}$ is the volume of a geometric tetrahedron with edge lengths $\lambda k_{ab}+\frac{1}{2}$ and $\Theta_{ab}$ are the dihedral angles.  Note that the formula scales as $\lambda^{-3/2}$ due to the volume term.
Currently, our formula for the tetrahedron contains the Regge action but the amplitude, phase term and scaling do not obviously agree with \eqref{pr formula}.
We will first consider the intertwiner normalisation, which will be necessary to obtain the correct scaling behaviour and some numerical factors, and then evaluate the Hessian numerically to check the agreement of the remaining terms.  The main drawback of the coherent state approach occurs here as it is very difficult to obtain an analytic formula for the determinant of the Hessian matrix.

\subsubsection{Intertwiner normalisation}
For the 6j symbol, the three valent intertwiners are normalised by dividing by the square root of the theta spin network.  The coherent intertwiners that we replaced these with, however, are not normalised.  The normalisation of these intertwiners for the coherent tetrahedron was studied in \cite{livine-2007-76} so we will briefly summarise the results for the case of the coherent triangle.

The normalisation of the coherent intertwiner is given in terms of the three edge vectors of the triangle $\nb_1 , \nb_2 ,\nb_3$ by the Hermitian inner product
\bea
f_{\Delta}( \nb_i,k_i)
&=& \int_{\SU(2)} dX  \prod_{i=1}^3 \langle \nb_i , k_i |   X   | \nb_i , k_i \rangle
\nn \\
&=& \int_{\SU(2)} dX  \exp S_{\Delta}
\eea
where
\be
S_{\Delta} = \sum_{i=1}^3 2k_i \ln \langle \nb_i  |   X   | \nb_i \rangle.
\ee
This integral can be calculated exactly using \cite{perelomov,KauffmanLins}, the result being
\bea
f_{\Delta}
&=&
\frac{
(1- \nb_1 . \nb_2)^p
(1- \nb_1 . \nb_2)^q
(1- \nb_1 . \nb_2)^r
(p+q)!(q+r)!(p+r)!
}
{2^{p+q+r} (p+q+r+1)! p!q!r!  }.
\eea
Where $p=k_1+k_2 -k_3$, $q=k_2+k_3 -k_1$ and $r=k_1+k_3 -k_2$.

The asymptotics of this intertwiner normalisation can also be found using stationary phase \cite{livine-2007-76}.
The stationarity of the action $S_{\Delta}$ gives the closure condition and the action evaluated on the critical points $\pm I$ gives zero.
\bea
f_{\Delta}(\nb_i , \lambda k_i) &\sim &
\left( \frac{2\pi}{\lambda} \right)^{3/2}  \frac{2}{(4\pi)^2} \frac{1}{\sqrt{\det H_{\Delta}}} \nn \\
&=&
 \frac{1}{     \sqrt{ 2^3 \pi \lambda^3     \det H_{\Delta}   }    }
\eea
The additional factor 2 comes from the fact that both $I$ and $-I$ are critical points that give the same contribution to the action.  $H_{\Delta} $ is the Hessian matrix of the action which is given by
\bea
H_{\Delta}^{ij} &=& \frac{\partial^2 S_\Delta }{\partial X^i   \partial X^j} \nn \\
&=& \frac{1}{2}\sum_l   k_l  (\delta^{ij}  -  n_l^i  n_l^j   )
\eea

We can now normalise our formula such that it agrees with the standard normalisation by dividing by a factor $ ( f_{\Delta_a})^{1/2}$ for each triangle $a$.

\subsubsection{Numerical calculations}
With the intertwiner normalisations included in the asymptotic formula, we obtain
\bea
\label{6j formula}
 \left\lbrace
\begin{array}{ccc}
\lambda k_{12} & \lambda k_{13} & \lambda k_{14} \\
\lambda k_{23} & \lambda k_{24} &  \lambda k_{34}
\end{array}
\right\rbrace
& =&
\frac{\Z_{PR}(\Psi_\lambda , \sigma  )}{   \prod_{p=1}^4  \sqrt{f_{\Delta_p}}    } \nn  \\
  &= &
\left( \frac{2\pi}{\lambda} \right)^{9/2}  \frac{   2^4    }{((4\pi)^2)^3   \sqrt{|\det H|}         } \frac{ 1   }{\prod_{p=1}^4  \sqrt{f_{\Delta_p}} }
\nn \\
&&
\times \cos \left( \sum_{a<b} \lambda k_{ab} \Theta_{ab}  - \frac{1}{2} \mathrm{Arg}(\det H)  \right)
\eea
Note that we have the correct scaling behaviour once the additional scaling factors from the intertwiners are included.  The normalisation terms are real so do not contribute any additional phase.

\begin{figure}
\begin{center}
\includegraphics[scale=.5]{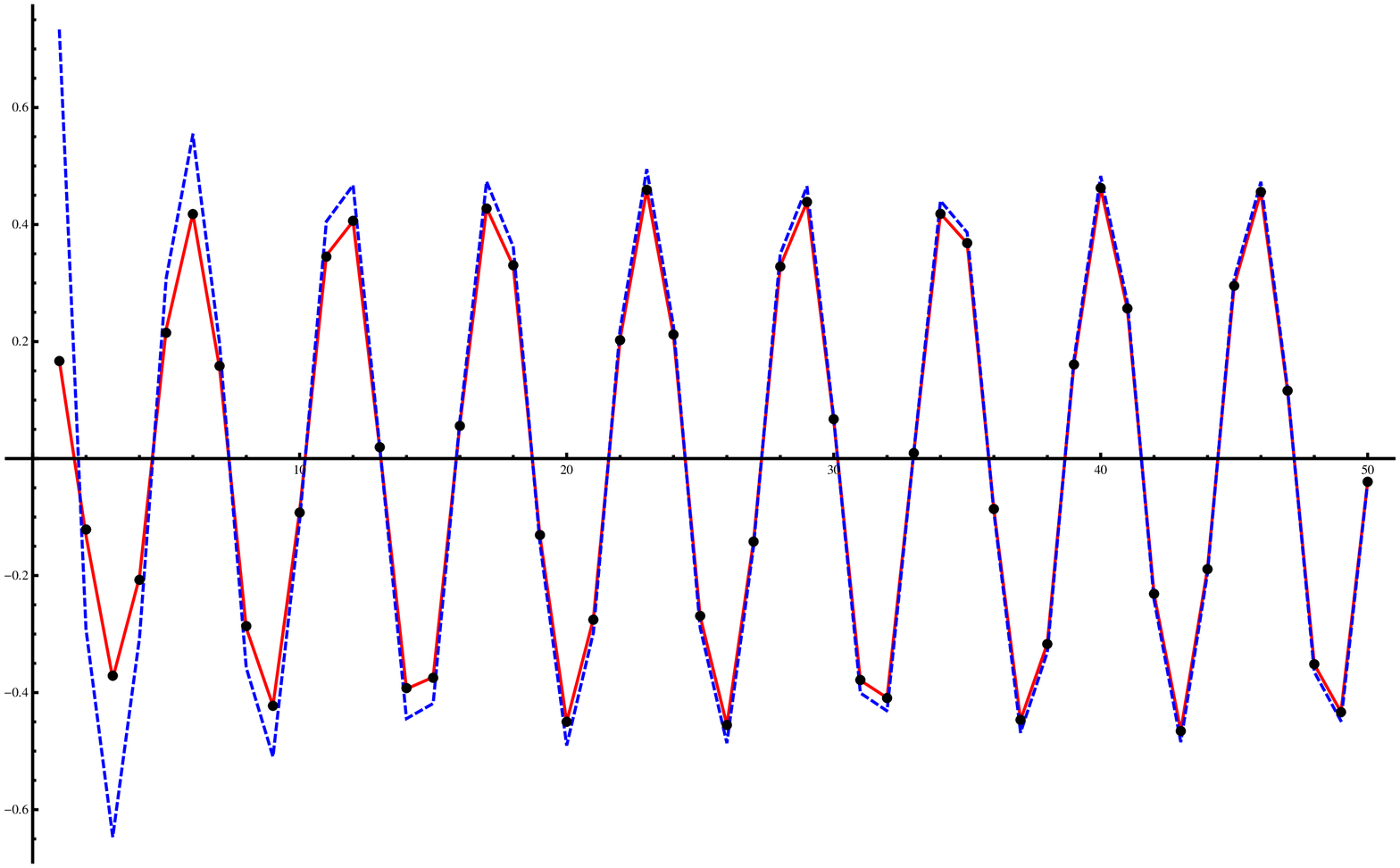}
\caption{Comparison of the 6j symbol (dots), the PR formula (red line) and equation \eqref{6j formula} (blue dashed line) against the scaling $\lambda$.  The scaling factor $\lambda^{-3/2}$ has been removed to make the comparison easier at low spins.}
\label{pr_vs_dgh plot}
\end{center}
\end{figure}

The formula for the equilateral tetrahedron with both the exact and approximate intertwiner normalisation was compared to the 6j symbol and the Ponzano-Regge asymptotic formula using Mathematica in Figure \ref{pr_vs_dgh plot}. We see that our formula differs from the Ponzano Regge formula for low spins. The only point where our formula differs from Ponzano Regge is in the fact that the Ponzano Regge asymptotics are given in terms of the dihedral angles and volume of the tetrahedron with edge lengths $\lambda k_{ab} + \frac{1}{2}$. Therefore the dihedral angles and volume change nontrivially with $\lambda$. A stationary phase approximation extracts only the scaling behaviour with respect to lambda in the asymptotic regime and cannot register this type of low spin behaviour.
%The difference between the 6j symbol and the asymptotic formula for the equilateral tetrahedron is plotted in Figure \ref{Z - 6j plot}.
%\begin{figure}
%\begin{center}
%\includegraphics[scale=0.4]{Zminus6j-0-200}
%\caption{Plot of the absolute value of the difference between the 6j symbol and the asymptotic formula against $\lambda$.}
%\label{Z - 6j plot}
%\end{center}
%\end{figure}
%We also compare this formula to the Ponzano-Regge asymptotic formula in Figure \ref{Z - 6j plot}.
This agrees as well as the PR formula for larger spins, however the agreement for very low values is not as good - Figure \ref{pr_vs_dgh plot}.

\section{Example: Steffen's flexible polyhedron}

\begin{figure}
\begin{center}
\psfrag{5}{$5$}
\psfrag{6}{$6$}
\psfrag{2.5}{$2.5$}
\psfrag{5.5}{$5.5$}
\psfrag{8.5}{$8.5$}
\includegraphics[scale=0.75]{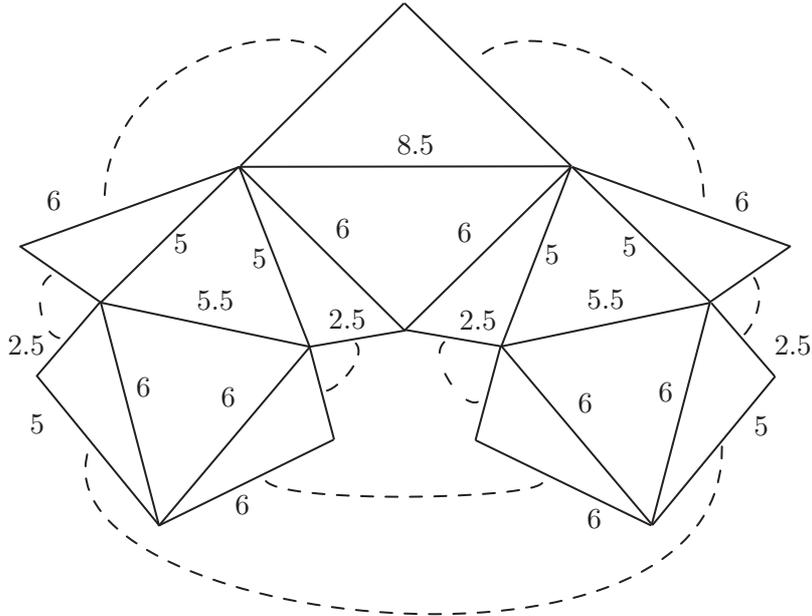}
\caption{A net showing a set of boundary data that reconstructs Steffen's flexible polyhedron.}
\label{steffdiagram}
\end{center}
\end{figure}

\begin{figure}
\begin{center}
\psfrag{j5}{$5$}
\psfrag{j6}{$6$}
\psfrag{j5/2}{$5/2$}
\psfrag{j10/2}{$10/2$}
\psfrag{j17/2}{$17/2$}
\includegraphics[scale=0.65]{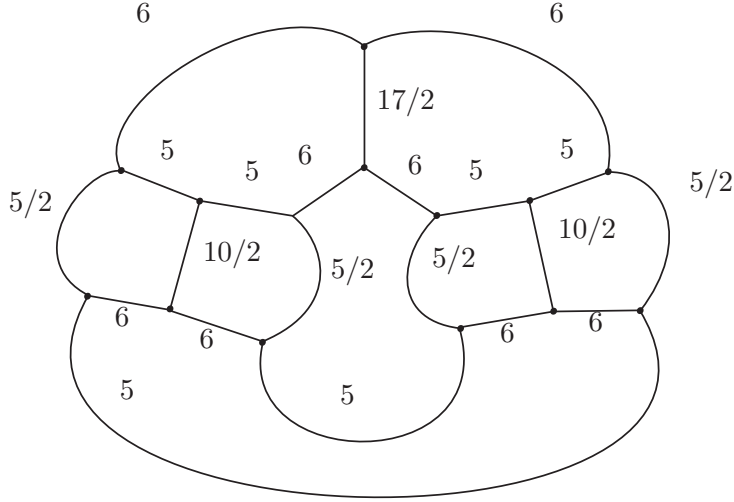}
\caption{The spin network corresponding to Steffen's flexible polyhedron.}
\label{steffnetwork}
\end{center}
\end{figure}

Here we discuss an example for which the second part of theorem \ref{big theorem new} is relevant, that is we describe a set of boundary data that admits a flexible immersion.
This particular example is taken from a flexible polyhedron with half integer edge lengths consisting of fourteen boundary triangles which was found by K. Steffen \cite{steffen}.
A net for constructing this polyhedron is given in Figure \ref{steffdiagram} and the corresponding spin network in Figure \ref{steffnetwork}.
Since Steffen's polyhedron admits a flex in one direction, we know that the flexifold is at least one dimensional.  As a polyhedron, it is not allowed to self intersect but there may be other immersions with flexibility in more than one dimension.
Applying the asymptotic formula with the same intertwiner normalisation as the tetrahedron in section \ref{tet section}, we would expect the scaling to be $\lambda^{-17/2}$.
%- see Figure \ref{steffbla}.
%\begin{figure}
%\begin{center}
%\includegraphics[scale=0.65]{Steffenbla}
%\caption{Scaling of the Steffen network compared to the expected result on a log plot.}
%\label{steffbla}
%\end{center}
%\end{figure}

\section{Discussion and Conclusions}

\subsection{Rigidity of cut immersions}

As discussed, the asymptotic formula produces a sum over all possible immersions of the boundary data in $\R^3$, including flexible ones.  These flexible immersions scale differently and thus dominate the rigid immersions. The question of whether a particular polyhedron is rigid is a difficult long standing problem in mathematics. A classic result is that convex polyhedra are in fact rigid, however this does not extend to immersions and non convex polyhedra where counter examples, like Steffen's polyhedron discussed above, are known.

If the boundary data is topologically $\mathcal{S}^2$ then a theorem by Steinitz \cite{gluck} applies that states that any simplicial complex with underlying space homeomorphic to a 2-sphere admits a simplexwise linear embedding into $\mathbb{R}^3$ whose image is strictly convex. This embedding will indeed be rigid and we can conclude that for the ball $\mathfrak{I}_r$ will always be non-empty.

To our knowledge the only more general results on rigidity of immersions are those giving conditions on bar frameworks, that is a graph immersed in $\mathbb{R}^d$, to be generically rigid. A bar framework is considered generic if the coordinates of the vertices are algebraically independent over the rationals, that is, there is no polynomial with rational coefficients that has these coordinates as roots. A graph is generically rigid if all its generic frameworks are. A set of sufficient and neccessary conditions for a graph to be generically rigid are known \cite{Connelly-Global, gortler-2007}. Unfortunately this does of course not cover our case with half-integer edge lengths.

Concerning the rigidity of cut immersions, which can be seen as bar frameworks with additional constraints, nothing is known.

\subsection{Surface immersions vs interior immersions}

With the asymptotic analysis performed above, we explicitly obtain a sum over immersions of the boundary data weighted by the cosine of the Regge action for the immersed surface.  Previously, asymptotics of the Ponzano-Regge model for larger triangulations could only be considered by taking the product of the asymptotic formula for each 6j symbol.  We now illustrate schematically that, in a simple example, that this is in fact equivalent to the asymptotic formula above.

\begin{figure}
\begin{center}
\psfrag{t}{$\Delta$}
\psfrag{s1}{$\sigma_1$}
\psfrag{Ps2}{$P \sigma_2$}
\psfrag{s2}{$\sigma_2$}
\includegraphics[scale=0.3]{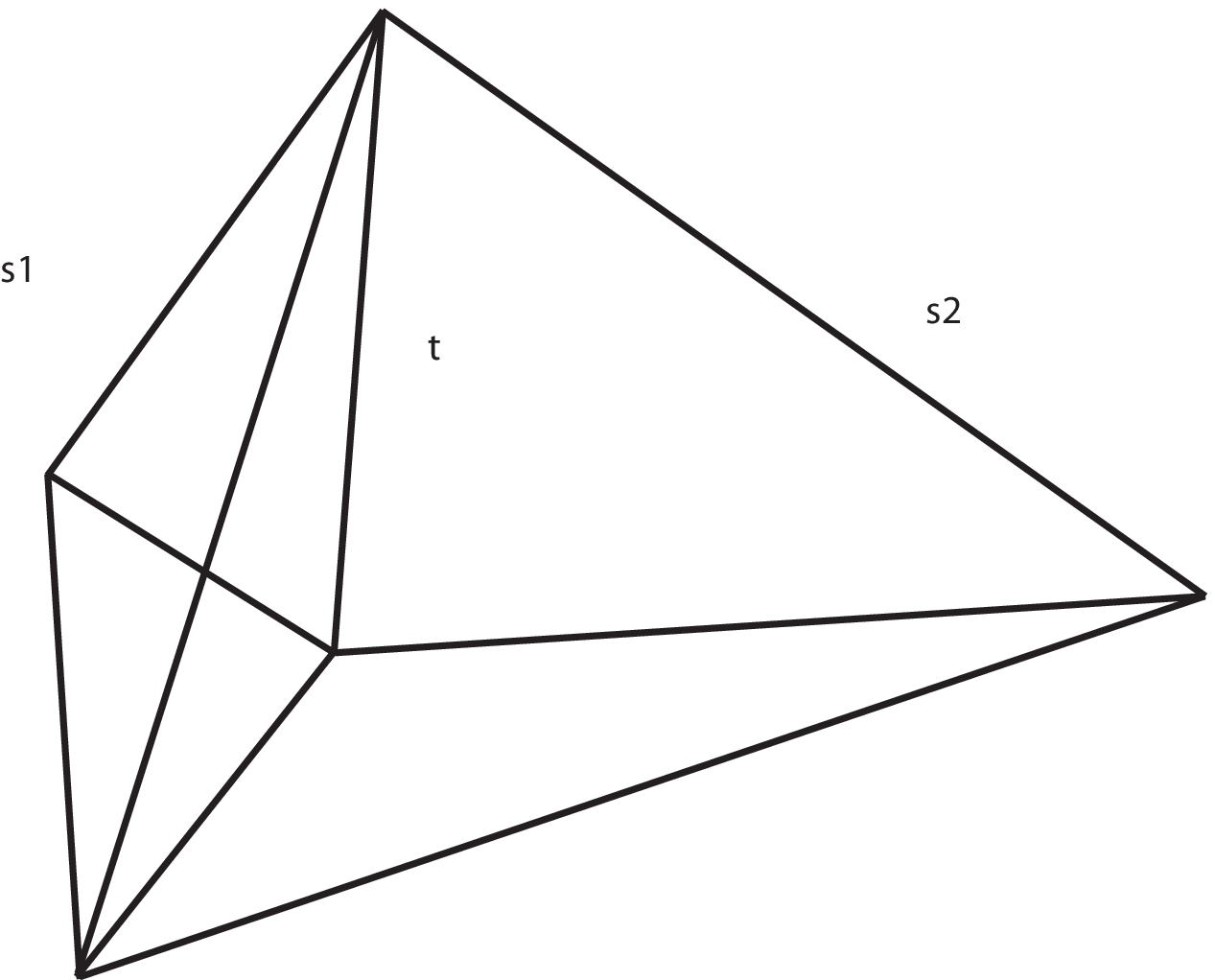} \hspace{10mm}
\includegraphics[scale=0.3]{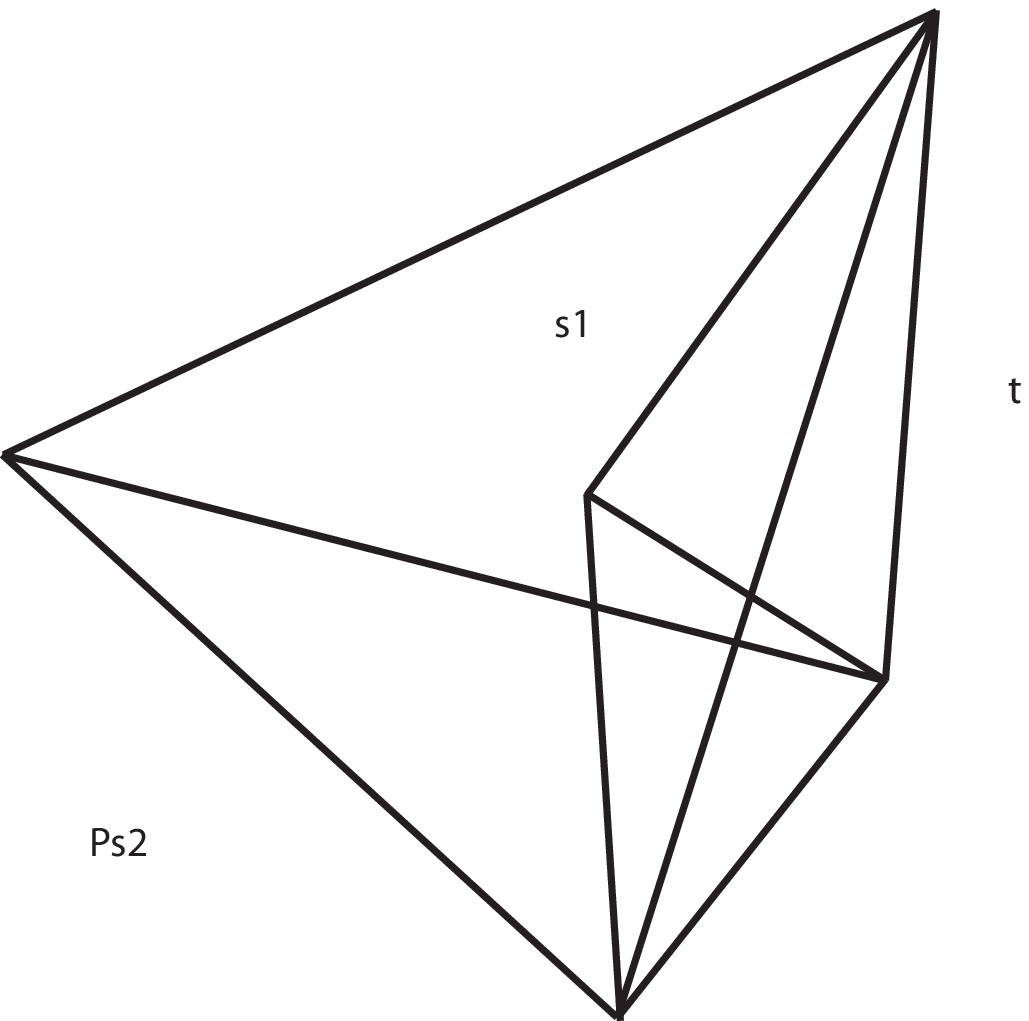}
\caption{Two different possible immersions of the boundary data for two tetrahedra $\sigma_1 , \sigma_2$ glued on a common triangle $\Delta$.}
\label{immersions}
\end{center}
\end{figure}

We will consider the case of two tetrahedra $\sigma_1 , \sigma_2$ glued along a common triangle $\Delta$ and use the boundary normalisation that agrees with the 6j symbol.  The partition function then reads
\be
\Z_{PR}(\Psi_\lambda , \sigma_1  \cup_\Delta   \sigma_2 ) =
 \left\lbrace
\begin{array}{ccc}
\lambda k_{1} & \lambda k_{2} & \lambda k_{3} \\
\lambda k_{4} & \lambda k_{5} &  \lambda k_{6}
\end{array}
\right\rbrace
%%%%%%%%%%%%%%%%%%%%%%%%%%%%
\left\lbrace
\begin{array}{ccc}
\lambda k_{1} & \lambda k_{2} & \lambda k_{3} \\
\lambda k_{7} & \lambda k_{8} &  \lambda k_{9}
\end{array}
\right\rbrace
\ee
We write the asymptotic formula for the 6j in terms of the Regge action $S_\sigma$ for a tetrahedron $\sigma$
\be
\label{simple pr formula}
 \left\lbrace
\begin{array}{ccc}
\lambda k_{1} & \lambda k_{2} & \lambda k_{3} \\
\lambda k_{4} & \lambda k_{5} &  \lambda k_{6}
\end{array}
\right\rbrace
=
N  \left(  \exp ( i\lambda S_{\sigma}) + \exp (- i\lambda S_{\sigma})  \right)
\ee
Where several of the factors have been absorbed into the amplitude $N$.  Asymptotically, this gives
\bea
\Z_{PR}(\Psi_\lambda , \sigma_1  \cup_\Delta   \sigma_2 )
&=& N_1 N_2  \left(  \exp ( i \lambda (S_{\sigma_1} +S_{\sigma_2}     ) ) + \exp (- i\lambda (S_{\sigma_1}+S_{\sigma_2} ) \right)
\nn \\
&&+ N_1 N_2  \left(  \exp ( i \lambda (S_{\sigma_1} - S_{\sigma_2}     ) ) + \exp (- i\lambda (S_{\sigma_1}-S_{\sigma_2} ) \right)
\nn \\
&=& N_1 N_2  \exp (\sum_{e \subset \Delta} k_e \pi)  \left(    \cos \left(    \lambda (S_{\sigma_1 \cup_t \sigma_2}   \right)
+    \cos \left(    \lambda (S_{\sigma_1 \cup_\Delta P \sigma_2}   \right) \right)
\eea
Where $P\sigma$ is the parity related tetrahedron and we have used the fact that the Regge action for two tetrahedra becomes
\be
S_{\sigma_1} + S_{\sigma_1} = S_{\sigma_1 \cup_\Delta \sigma_2}  + \sum_{e \subset \Delta} k_e \pi.
\ee
Thus the formula gives a sum over the two different ways of immersing the boundary triangles in $\mathbb{R}^3$, see Figure \ref{immersions} .
If we now consider larger triangulations, possibly with interior vertices, edges or faces, then the immersed boundary surfaces we obtain are related to immersions of the entire triangulation by adding the immersions of the interior triangulation. Note that the internal vertices can be placed anywhere in $\R^3$ which leads to the divergent factors that appear in the Ponzano-Regge state sum for these kind of triangulations. The two parts of the cosine appear because these immersions can be orientation reversing on particular tetrahedra. These will then contribute with negative dihedral angles to the interior action. Thus the integration over possible interior geometries can be split into parts corresponding to particular orientations on the interior tetrahedra. The terms of this sum will then correspond to the sum of terms obtained by multiplying out the sums in the interior cosines. On shell the interior action on these immersions is zero so the overall amplitude only registers the boundary contribution to the action. In particular interior configurations corresponding to different geometries and orientations on the interior, and thus to different terms in the cosine, contribute with the same phase.

\subsection{Boundary states}

We also note that it is possible to select a particular immersion in the sum by choosing a boundary state peaked around a particular set of dihedral angles, see for example \cite{Rovelli:2005yj,Speziale:2005ma,Bianchi:2009ri}.  This boundary state also selects one overall orientation of the immersion which removes the parity related term in the asymptotic formula.  For non-rigid immersions, the boundary state would also have the ability to select a particular configuration of the immersed surface which would stop these immersions dominating the integral.

A possible problem with the boundary state is that while it selects an orientation for the boundary, it was not clear if the orientations of the interior tetrahedra behaved consistently.  This was considered in \cite{Bianchi:2008ae} and our result also suggests that these do not cause a problem as the asymptotic formula does not register these orientations.

\subsection{Conclusions}

In this paper we addressed the problem of asymptotics of larger triangulations for the Ponzano-Regge model.  By reformulating the partition function as a spin network on the boundary and then rewrote this amplitude using $\SU(2)$ coherent states. While this particular feature will not be available for non topological theories one could expect that in general boundary data will be not suppressed if it can be continued to a solution of the equations of motion on the interior. The asymptotic formula contains a sum over immersions of the boundary data weighted by the cosine of the Regge action. Interestingly, Ponzano and Regge point out in \cite{ponzanoregge} that the different possible immersions corresponding to 3-nj symbols should contribute to the asymptotics but did not obtain a concrete formula.
The presented work opens up the possibility to do an exhaustive analysis of the classical limit of the Ponzano Regge model including correlation functions on the boundary. As such it can serve as a toy model and proof of concept for conceptual issues likely to arise in all background independent theories.

Of further interest would be to consider in more detail how the asymptotics obtained here can be obtained from the ``product of cosines'' picture. In particular to shed light on the issue of causality and orientation in spin foam models.

Interestingly, and unexpectedly, we found that spin networks contain some information about the rigidity properties of surfaces. The scaling properties of a spin network correspond to the maximum dimension of flexibility if the geometry to which it corresponds has any non-rigid immersions.

\section{Acknowledgements}

RD and FH are funded by EPSRC doctoral grants. We would like to thank John Barrett for discussions and comments on a draft of this paper.

%%%%%%%%%%%%%%%%%%%%%%%%%%%%%%%%%%%%%%%%%%%%%%%%%%%%%%%%%%%%%%%%%%%%%%%%%%%%%%%%%%

%%%%%%%%%%%%%%%%%%%%%%%%%%%%%%%%%%%%%%%%%%%%%%%%%%%%%%%%%%%%%%%%%%%%%%%%%%%%%%%%%%
\appendix
\section{Example of the Ponzano-Regge amplitude as a spin network on the boundary of the solid torus }
\label{tet appendix}
Here we give a simple example of Lemma \ref{lemma2} on the solid torus $\mathbb{T}$.  A non-tardis (degenerate) triangulation of the solid torus with three tetrahedra is given by
\begin{center}
\psfrag{a}{$k_1$}
\psfrag{b}{$k_2$}
\psfrag{c}{$k_3$}
\psfrag{d}{$k_4$}
\psfrag{e}{$k_5$}
\psfrag{f}{$k_6$}
\psfrag{g}{$k_7$}
\psfrag{h}{$k_8$}
\psfrag{i}{$k_9$}
\includegraphics[scale=0.3]{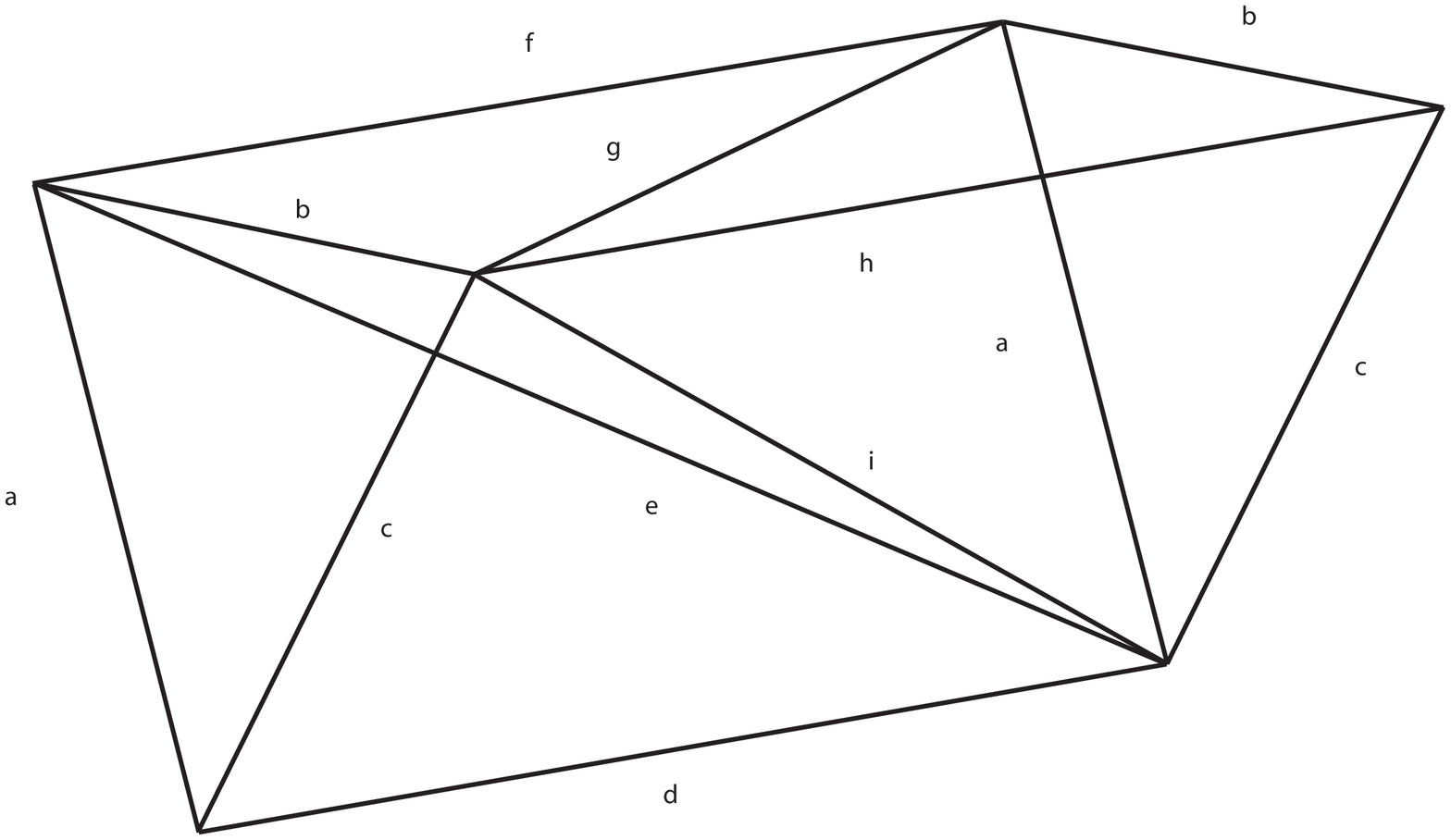}
\end{center}
The two triangles with edges $k_1,  k_2 , k_3$ are identified.
The Ponzano-Regge amplitude is given by
\be
\label{PR torus example}
\mathcal{Z}_{PR}(\Psi, \mathbb{T})
=
\left\{
\begin{array}{ccc}
 k_1 & k_2 & k_3 \\
 k_8 & k_9 & k_7
\end{array}
\right\}
\left\{
\begin{array}{ccc}
 k_1 & k_2 & k_3 \\
 k_9 & k_4 & k_5
\end{array}
\right\}
\left\{
\begin{array}{ccc}
 k_1 & k_5 & k_6 \\
 k_2 & k_7 & k_9
\end{array}
\right\}.
\ee
We choose the cutting disc $\mathbb{D}$ to be the triangle $k_1, k_2, k_3$ and perform the cut that reduces $\mathbb{T}$ to the 3-ball.
A net for constructing the triangulation on the boundary is given by
\begin{center}
\psfrag{a}{$k_1$}
\psfrag{b}{$k_2$}
\psfrag{c}{$k_3$}
\psfrag{d}{$k_4$}
\psfrag{e}{$k_5$}
\psfrag{f}{$k_6$}
\psfrag{g}{$k_7$}
\psfrag{h}{$k_8$}
\psfrag{i}{$k_9$}
\includegraphics[scale=0.25]{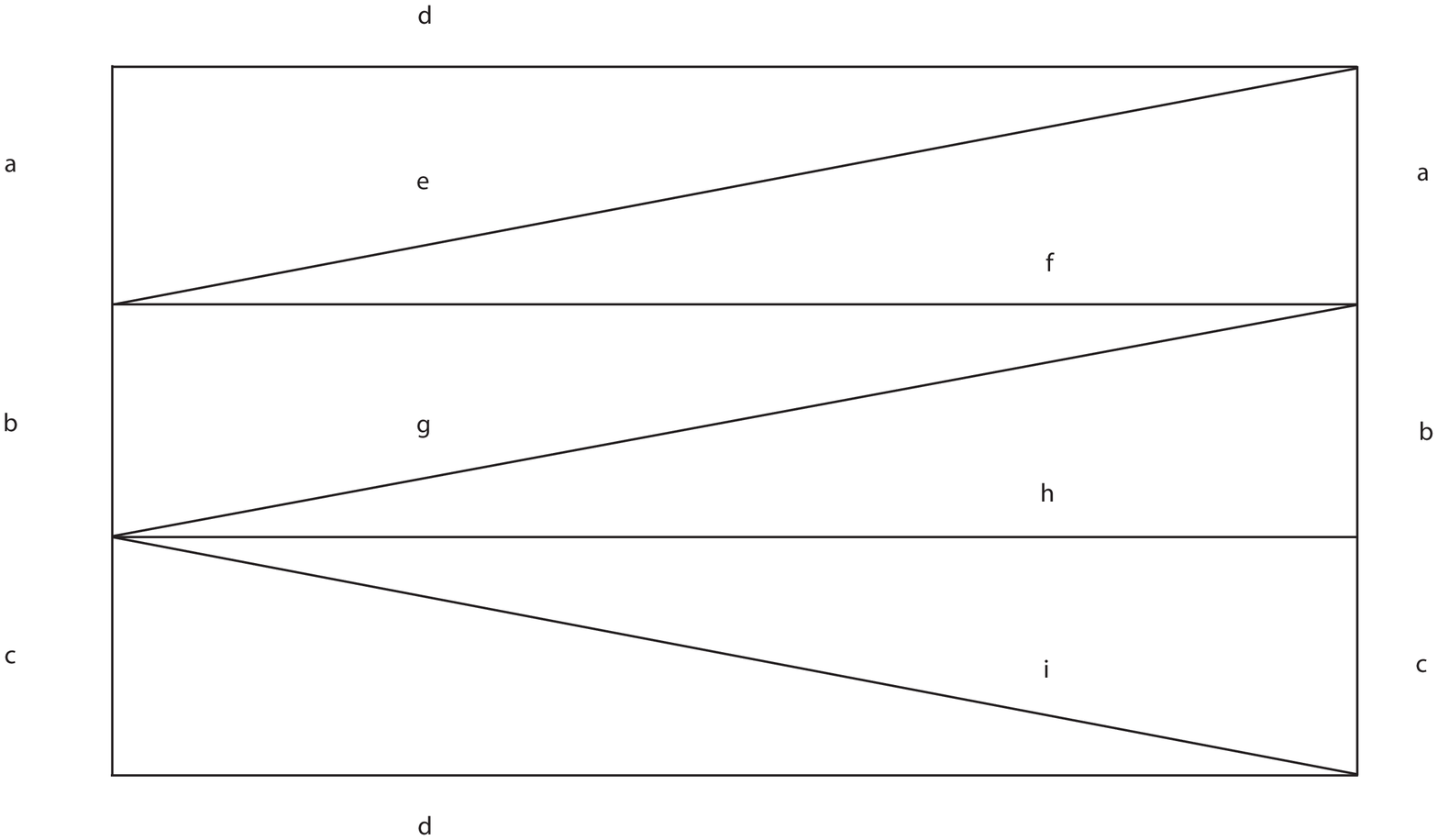}
\end{center}
The Ponzano-Regge amplitude can be expressed as the following spin network evaluation on the boundary, with a group integral inserted on each of the dual edges that cross $\mathbb{D}$.
\begin{center}
\psfrag{a}{$k_1$}
\psfrag{b}{$k_2$}
\psfrag{c}{$k_3$}
\psfrag{d}{$k_4$}
\psfrag{e}{$k_5$}
\psfrag{f}{$k_6$}
\psfrag{g}{$k_7$}
\psfrag{h}{$k_8$}
\psfrag{i}{$k_9$}
\psfrag{H}{$h$}
\psfrag{int}{$\mathcal{Z}_{PR}(\Psi, \mathbb{T}) = \int_{\SU(2)} dh$}
\includegraphics[scale=0.35]{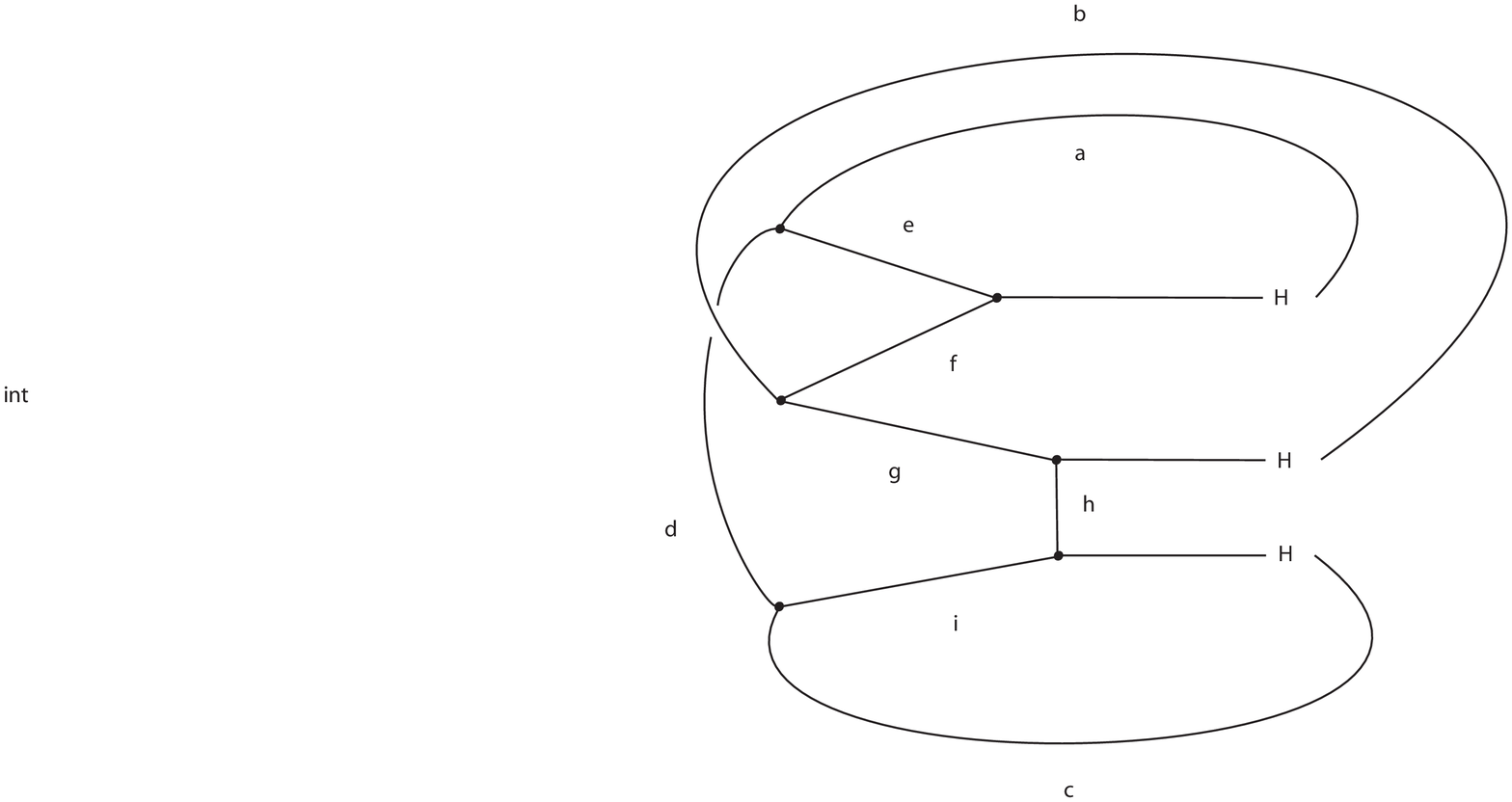}
\end{center}
Expressing this spin network in terms of 6j symbols gives equation \eqref{PR torus example}.

%%%%%%%%%%%%%%%%%%%%%%%%%%%%%%%%%%%%%%%%%%%%%%%%%%%%%%%%%%%%%%%%%%%%%%%%%%%%%%%%%%%%%%%%%%%%%%%%%%%%%%%%%%%%%%%%%%%%%%%%%%%%%%%%%

%%%%%%%%%%%%%%%%%%%%%%%%%%%%%%%%%%%%%%%%%%%%%%%%%%%%%%%%%%%%%%%%%%%%%%%%%%%%%%%%%%%%%%%%%%%%%%%%%%%%%%%%%%%%%%%%%%%%%%%%%%%%%%%%%

%%%%%%%%%%%%%%%%%%%%%%%%%%%%%%%%%%%%%%%%%%%%%%%%%%%%%%%%%%%%%%%%%%%%%%%%%%%%%%%%%%%%%%%%%%%%%%%%%%%%%%%%%%%%%%%%%%%%%%%%%%%%%%%%%

%%%%%%%%%%%%%%%%%%%%%%%%%%%%%%%%%%%%%%%%%%%%%%%%%%%%%%%%%%%%%%%%%%%%%%%%%%%%%%%%%%%%%%%%%%%%%%%%%%%%%%%%%%%%%%%%%%%%%%%%%%%%%%%%%

%%%%%%%%%%%%%%%%%%%%%%%%%%%%%%%%%%%%%%%%%%%%%%%%%%%%%%%%%%%%%%%%%%%%%%%%%%%%%%%%%%%%%%%%%%%%%%%%%%%%%%%%%%%%%%%%%%%%%%%%%%%%%%%%%

\bibliographystyle{hieeetr}
\bibliography{Bibliography2009}

\begin{thebibliography}{10}

\bibitem{ponzanoregge}
G.~Ponzano and T.~Regge, ``Semiclassical limit of racah coefficients,'' in {\em
  Spectroscopy and group theoretical methods in physics} (F.~Block, ed.),
  pp.~1--58, North Holland, 1968.

\bibitem{roberts-1999-3}
J.~Roberts, ``Classical 6j-symbols and the tetrahedron,'' {\em GEOM.TOPOL.},
  vol.~3, p.~21, 1999.

\bibitem{gurau-2008}
R.~Gurau, ``The ponzano-regge asymptotic of the 6j symbol: an elementary
  proof,'' 2008.

\bibitem{garoufalidis-2009}
S.~Garoufalidis and R.~{van der Veen}, ``Asymptotics of classical spin
  networks,'' 2009, 0902.3113.

\bibitem{barrett-2003-20}
J.~W. Barrett and C.~M. Steele, ``Asymptotics of relativistic spin networks,''
  {\em Classical and Quantum Gravity}, vol.~20, p.~1341, 2003.

\bibitem{Freidel:2002mj}
L.~Freidel and D.~Louapre, ``{Asymptotics of 6j and 10j symbols},'' {\em Class.
  Quant. Grav.}, vol.~20, pp.~1267--1294, 2003, hep-th/0209134.

\bibitem{Dupuis:2009sz}
M.~Dupuis and E.~R. Livine, ``{Pushing Further the Asymptotics of the
  6j-symbol},'' {\em Phys. Rev.}, vol.~D80, p.~024035, 2009, 0905.4188.

\bibitem{Barrett:2008wh}
J.~W. Barrett and I.~Naish-Guzman, ``{The Ponzano-Regge model},'' {\em Class.
  Quant. Grav.}, vol.~26, p.~155014, 2009, 0803.3319.

\bibitem{Barrett:2009gg}
J.~W. Barrett, R.~J. Dowdall, W.~J. Fairbairn, H.~Gomes, and F.~Hellmann,
  ``{Asymptotic analysis of the EPRL four-simplex amplitude},'' 2009,
  0902.1170.

\bibitem{Barrett:2009mw}
J.~W. Barrett, R.~J. Dowdall, W.~J. Fairbairn, F.~Hellmann, and R.~Pereira,
  ``{Lorentzian spin foam amplitudes: graphical calculus and asymptotics},''
  2009, 0907.2440.

\bibitem{Freidel:2007py}
L.~Freidel and K.~Krasnov, ``{A New Spin Foam Model for 4d Gravity},'' {\em
  Class. Quant. Grav.}, vol.~25, p.~125018, 2008, 0708.1595.

\bibitem{Engle:2007wy}
J.~Engle, E.~Livine, R.~Pereira, and C.~Rovelli, ``{LQG vertex with finite
  Immirzi parameter},'' {\em Nucl. Phys.}, vol.~B799, pp.~136--149, 2008,
  0711.0146.

\bibitem{livine-2007-76}
E.~R. Livine and S.~Speziale, ``A new spinfoam vertex for quantum gravity,''
  {\em Physical Review D}, vol.~76, p.~084028, 2007.

\bibitem{Speziale:2005ma}
S.~Speziale, ``{Towards the graviton from spinfoams: The 3d toy model},'' {\em
  JHEP}, vol.~05, p.~039, 2006, gr-qc/0512102.

\bibitem{Moussouris}
J.~Moussouris, {\em Quantum models of space-time based on recoupling theory}.
\newblock PhD thesis, St. Cross College, Oxford, 1983.

\bibitem{perelomov}
A.~Perelomov, {\em Generalized coherent states and their applications}.
\newblock Springer-Verlag, 1986.

\bibitem{KauffmanLins}
L.~H. Kauffman and S.~L. Lins, {\em Temperley-Lieb recoupling theory and
  invariants of 3-manifolds}.
\newblock Princeton University Press, 1994.

\bibitem{Hormander}
L.~Hormander, {\em The analysis of linear partial differential operators I}.
\newblock Springer-Verlag, 1983.

\bibitem{ramacher-2009}
P.~Ramacher, ``Singular equivariant asymptotics and the moment map i,'' 2009,
  0902.1248v1.

\bibitem{steffen}
{K. Steffen}, ``{A symmetric flexible Connelly sphere with only nine
  vertices}.'' {See:
  http://demonstrations.wolfram.com/SteffensFlexiblePolyhedron/ for Mathematica
  code}.

\bibitem{gluck}
H.~Gluck, ``Almost all simply connected surfaces are rigid.,'' in {\em Lecture
  notes in mathematics 438: Geometric Topology}, Springer-Verlag, 1975.

\bibitem{Connelly-Global}
R.~Connelly, ``Generic global rigidity,'' {\em Discrete Comput. Geom.},
  vol.~33, no.~4, pp.~549--563, 2005.

\bibitem{gortler-2007}
S.~J. Gortler, A.~D. Healy, and D.~P. Thurston, ``Characterizing generic global
  rigidity,'' 2007, 0710.0926.

\bibitem{Rovelli:2005yj}
C.~Rovelli, ``{Graviton propagator from background-independent quantum
  gravity},'' {\em Phys. Rev. Lett.}, vol.~97, p.~151301, 2006, gr-qc/0508124.

\bibitem{Bianchi:2009ri}
E.~Bianchi, E.~Magliaro, and C.~Perini, ``{LQG propagator from the new spin
  foams},'' 2009, 0905.4082.

\bibitem{Bianchi:2008ae}
E.~Bianchi and A.~Satz, ``{Semiclassical regime of Regge calculus and spin
  foams},'' {\em Nucl. Phys.}, vol.~B808, pp.~546--568, 2009, 0808.1107.

\end{thebibliography}

\end{document}